\documentclass[trackchanges,twocolappendix,twocolumn,tighten]{aastex7}

\usepackage{multirow}
\usepackage{amsmath}
\usepackage{booktabs}
\usepackage{float}
\usepackage{array}

\begin{document}

\title{CLARA: A Modular Framework for Unsupervised Transit Detection Using TESS Light Curves}

\author[0009-0004-0973-0226]{Mainak Dasgupta}
\affiliation{Independent Researcher}
\email{mdasgupta0112@gmail.com}

\begin{abstract}

We present \textbf{CLARA}, a modular framework for unsupervised transit detection in TESS light curves, leveraging Unsupervised Random Forests (URFs) trained on synthetic datasets and guided by morphological similarity analysis.

\vspace{0.5em}
This work addresses two core questions: \newline

\noindent(a) How does the design of synthetic training sets affect the performance and generalization of URFs across independent TESS sectors?\\
\noindent(b) Do URF anomaly scores correlate with genuine astrophysical phenomena, enabling effective identification and clustering of transit-like signals?

\vspace{0.5em}
We investigate these questions through a two-part study focused on (1) detection performance optimization, and (2) the physical interpretability of anomalies.

\vspace{0.5em}
\textbf{In Part I}, we introduce three URF model variants tuned via $\alpha$-controlled scoring objectives, and evaluate their generalization across five TESS sectors. This large-scale test involved scoring 384,000 individual light curves (128,000 light curves per $\alpha$ variant), revealing stable, interpretable differences between recall-optimized, precision-optimized, and balanced models.

\vspace{0.5em}
\textbf{In Part II}, our optimized clustering (DPMM Cluster 2) yields a 14.04\% detection rate (16 confirmed transits among 114 candidates) from the first five TESS SPOC sectors. This reflects a substantial enrichment over baseline rates: 0.4569\% for the full TESS-SPOC project candidate set (7658 candidates across 1.68 million light curves), and 0.2650\% for the FFI-based SPOC sample (7658 candidates across 2.89 million light curves;). Additionally, we perform Monte Carlo injection-recovery tests to assess feasibility, stability, justification and quantification of methods outlined.

\vspace{0.5em}
All computations were performed on a personal, CPU-only desktop with an Intel\textsuperscript{\textregistered} Core\texttrademark~i3-8100 processor and 32~GB RAM, using parallelized scoring and classification routines across four physical cores (see Appendix \ref{sec:computeresourcesandruntime}). CLARA processed over 87,000 TESS SPOC light curves (Sectors 1–5) without GPU acceleration. The complete codebase is available on \href{https://github.com/googleboy-byte/CLARA}{Github}.

\end{abstract}




\section{Introduction}

The discovery and characterization of astrophysical phenomena from large-scale time-series surveys represents one of the most challenging problems in modern astronomy. With missions like the Transiting Exoplanet Survey Satellite (TESS), Kepler, K2 and the upcoming LSST observatory generating millions of light curves, traditional supervised approaches for anomaly detection face fundamental limitations due to the rarity and diversity of interesting astrophysical events . The need for scalable, unsupervised methods that can systematically discover and interpret anomalies - such as exoplanet transits, eclipsing binaries, and novel variable stars - has become increasingly urgent as survey data volumes continue to grow exponentially.

Recent advances in unsupervised machine learning for astronomical time-series analysis have shown promising results. \citet{Nun2015} demonstrated the effectiveness of Random Forest-based approaches for stellar variability classification, while \citet{Malanchev2021} explored deep learning methods for supernova discovery in large surveys. Active learning approaches have also emerged as powerful tools for personalized anomaly detection, with \citet{Lochner2016} introducing frameworks that combine human intuition with machine learning efficiency. The application of Unsupervised Random Forests (URFs) to exoplanet detection was demonstrated by \citet{Crake2023} (MG23 hereafter), who showed that synthetic training sets could guide anomaly detection behavior in TESS light curves, with an approach optimizing models for the sharpest anomaly score tail. From MG23: 
\begin{quote}
    Once the parameters that maximise this accuracy are identified, we fine-tune the specific hyperparameters near these values to increase the contrast in anomaly scores between objects.
\end{quote} 
However, these approaches have lacked systematic frameworks for controlling the anomaly discovery process and interpreting the physical nature of detected anomalies.

The challenge of anomaly detection in astrophysical time-series data is compounded by several factors: the high dimensionality of light curve features, the presence of various noise sources and systematic effects, and the need to distinguish between different classes of astrophysical phenomena \citet{Ivezic2019}. Traditional clustering and dimensionality reduction techniques often fail to provide physically meaningful groupings of astronomical objects, while purely data-driven approaches may discover statistically significant but astrophysically uninteresting patterns and often need large sets of training data for targetted discovery. Then there are physics driven algorithms like Box Least Squares and Transit-Least-Squares which provide fixed detection strategies. \newline

We present CLARA (Controllable Learning for Anomaly Recognition in Astrophysics), a comprehensive unsupervised random forest model design pipeline that addresses these limitations through systematic synthetic set design and design-feature-to-performance-metric mapping. Our approach introduces novel methodologies for steering URF behavior through controlled variation of synthetic set generation parameters, enabling the targeted discovery of specific classes of astrophysical phenomena characterized by light curve morphology. \textbf{This is an unsupervised anomaly discovery algorithm with controllable behavior and semi-supervised downstream filtering for astrophysical contextualization of anomalies found.} Furthermore, we develop morphological classification techniques using cosine similarity matching with known objects and integrate astrometric data from Gaia DR3 \citet{GaiaCollaboration2022} to provide physical interpretation of discovered anomalies. This work aims to transform unsupervised anomaly detection from a purely statistical exercise into a physically guided discovery and prioritization tool for astronomical surveys.

This paper is organized into two main parts, corresponding to the two research questions outlined in the abstract. \newline

\textbf{Part I} investigates how synthetic light curve construction for fine tuning of hyper-parameters influences the performance of Unsupervised Random Forest (URF) models for anomaly detection in TESS 2-minute cadence SPOC light curves. It introduces a parameterized scoring metric controlled by a variable coefficient $\alpha$ that allows balancing of detection recall and astrophysical significance. We perform a grid search over synthetic set parameters (e.g., duration, noise level, cadence) and predict model behavior using TOI (Tess Object of Interest) Recall, Importance metrics for TOIs recovered by URF models, Anomaly Rate and other performance metrics over a small stratified test set. Then we establish generalization of the predicted model performance across both similar subsets from the same sector and diverse populations across five different sectors, presented later. \newline

\textbf{Part II} validates the astrophysical significance of high-scoring URF anomalies. We introduce a revised scoring scheme (Normalized Weighted Root Sum of Squares, or N-WRSS), and correlate anomaly scores with known stellar parameters (e.g., $T_{\mathrm{eff}}$, $\log g$, RUWE, $v_{\mathrm{tan}}$) from Gaia DR3. Morphological classification is performed using cosine similarity to small representative set of labeled TOI features with SIMBAD object types (e.g., Pl, EB*, V*, PM*), allowing us to quantify the physical plausibility of detected signals. Then we finish by introducing an advance 10 dimensional feature set for enhanced clustering of transit-like curves to filter out non-transit-like curves.\newline

\textbf{A key point to note is that rather than competing with established transit detection methods, CLARA demonstrates how unsupervised ML frameworks can be systematically tuned and validated for specific discovery objectives using synthetically generated data - a capability with broad applications across astronomical time-domain surveys, including prioritization of targets when faced with limited follow-up resources, identification of rare or anomalous events in large datasets that have high probability of fitting an astrophysical class, and adaptive survey strategies that can evolve based on emerging scientific priorities or unexpected discoveries.}

\section{Methods}

\subsection{Part 1 - Exploring how the design of synthetic sets affects unsupervised anomaly detection}

Unsupervised random forest models for anomaly detection were first used in astronomical datasets in \citet{Baron2017} where isolation forests were applied to SDSS galaxy data to identify the most unusual and potentially rare objects in the survey. As demonstrated ahead, the controllability of unsupervised anomaly detection hinges on our ability to systematically design synthetic training sets that guide model behavior toward discovering specific types of astrophysical phenomena. By varying key parameters such as the number of synthetic curves in the artificially generated data set, duration of curves/transits, frequency, and noise characteristics in our synthetic light curves, we can predict the generalizable behavior of an URF model by testing on a small labeled representative dataset, allowing us to essentially \textbf{steer Unsupervised Random Forests} to prioritize different anomaly signatures and effectively "tune" the discovery process for targeted scientific objectives ranging from higher detection rate to better scoring of anomalies.

\subsection{Feature of a Curve - I}
\label{sec:curve_feature} 

The characteristic of a single light curve for input to our URF (Unsupervised Random Forest) model experiments follows established methods from our foundational paper \citet{Crake2023} where the first $n_{flux}$=3000 flux points are stacked on top of $n_{power}$=1000 Lomb-Scargle periodogram, \citet{VanderPlas2018} frequency points calculated using the LombScargle function of the astropy.timeseries module. The Lomb-Scargle periodogram is a method for detecting and estimating the strength of periodic signals in unevenly sampled time series data, extending the classical Fourier analysis to irregularly spaced observations as characteristic of TESS light curve data.  This function calculates the power of a period where the power describes the extent of the periodicity of the selected time value. Hence the feature is essentially a 4000 dimensional vector.
\begin{figure}
	\includegraphics[width=\columnwidth]{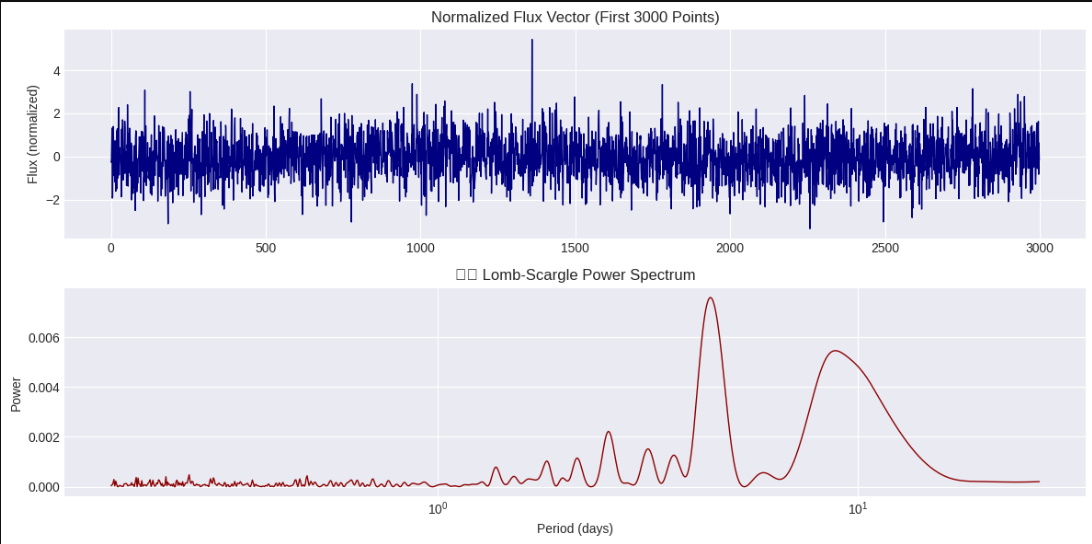}
    \caption{The first 3000 flux points of a light curve (top plot) and the Lomb-Scargle periodogram of the same curve (bottom plot)}
    \label{fig:feature_vec_sep_features_plot}
\end{figure}

\subsection{URF Hyperparamter Randomized Search via Real + Synthetic Curve Stacking}
\label{sec:urf_training_via_real_plus_synthetic_stacking}

For the real set (real features class), we took features (Section \ref{sec:curve_feature}) of 1500 real curves from TESS SPOC sector 2. A \verb|RandomizedSearchCV| is performed over a predefined hyperparameter space for the \verb|RandomForestClassifier|. The hyper-parameter search space is defined in Appendix \ref{sec:hyperparamsearchspaceappendix} \newline

For our initial tests, we took the following synthetic set (artificially generated features class) design types for the URF.
\newline
a. \textbf{URF1}. We generated the synthetic set by uniformly sampling feature values within the range of the real dataset as a baseline benchmark.
\newline
$X_{fake}$ = np.random.uniform(np.min($X_{real}$), np.max($X_{real}$), size=$X_{real}$.shape)
\newline
\newline
b. \textbf{URF2}. We used fixed hyperparameters from MG23 with n\_estimators=100, max\_features="sqrt", max\_depth=700, min\_samples\_split=4, min\_samples\_leaf=2, bootstrap=False, and warm\_start=True.
\newline
\newline
c. \textbf{URF3}. We used feature vectors from confirmed Tess Objects of Interest (TOIs) from Sector 1 of SPOC data as the synthetic set for hyperparameter search.
\newline
\newline
d. \textbf{URF4}. The features were computed from a set of synthetic light curves generated using transit models (box-shaped model and Mandel-Agol) injected into realistic noise baselines. The computational implementation of this was performed using the batman pypi package.
\newline
\newline
The four URF variants were designed to systematically explore different approaches to synthetic set construction for transit discovery. URF1 serves as a baseline control, using uniform random sampling within the feature space bounds to establish minimum performance expectations. URF2 tests the generalizability of the hyperparameters established in MG23, evaluating whether their optimization for stellar variability classification translates effectively to transit detection in TESS data. URF3 explores a realistic "ground truth" approach by using feature vectors from confirmed TOIs as the synthetic class, providing insight into the performance ceiling when the synthetic set closely matches actual transit signatures. Finally, URF4, our main URF variant for further work, was chosen because of \textbf{its inherent controllability over physically motivated synthetic dataset design, where synthetic light curves are generated using transit models with systematically variable parameters} (ratio of real curves - 1500 - to synthetic curves - 300), duration, cadence, and noise characteristics), enabling targeted optimization for specific discovery objectives and forming the foundation for our systematic parameter exploration. For URF4, we \textbf{fixed the real set to 1500 TESS SPOC curves across all experiments}; investigating how the size and composition of this real subset affects model performance is a valuable direction for future work. URF 1-4 model performance is as listed in Table \ref{tab:urf_benchmarkmodels_performance}.

\subsection{Anomaly Scoring}
\label{sec:anomalyscoringsubsection} 

Following \citet{Baron2017} and \citet{Crake2023}, our URF implementation uses terminal node population as the anomaly scoring heuristic. Each tree in the forest is trained to distinguish real light curves from synthetic ones, with anomaly scores computed by tracking the population composition of terminal nodes. For each real light curve, we calculate a similarity score S as the average fraction of real data points sharing the same terminal node across all trees in the forest. The final anomaly score is defined as 1 - S, where a score of 1 indicates maximum anomaly (the object consistently lands alone in terminal nodes) and 0 indicates minimal anomaly (the object is always grouped with the majority of real data). This population-based approach provides a natural measure of how isolated each light curve appears relative to the broader dataset distribution.

\subsection{Controlled Parameter Space Exploration}
\label{sec:controlledparamspaceexploration} 

URF-4's framework enables \textbf{systematic exploration of four key synthetic parameters} (and more not explored ones): 
\begin{enumerate}
    \item ncurves - the number of synthetic light curves in the artificially generated training set
    \item duration - temporal span of each synthetic curve in days
    \item cadence minutes - observational temporal resolution
    \item noise level - injected photometric noise in parts-per-million (ppm)

\end{enumerate}

The following parameter grid was used for this purpose:

\begin{verbatim}
    param_grid = {
    'n_curves': [100, 200, 300],
    'lc_length_days': [13.5, 27.0],
    'cadence_minutes': [2.0, 10.0],
    'noise_level': [50e-6, 100e-6, 200e-6],
    }
\end{verbatim}

Each parameter combination from the cartesian product space of the parameter grid values is used to fine tune hyper-parameters (\ref{sec:urf_training_via_real_plus_synthetic_stacking}) independent URF models against a fixed real light curve dataset, \textbf{resulting in \textbf{$3 \times 2 \times 2 \times 3 = 36$} sub-variant models.} 

\subsection{Evaluation Methodology}
\label{sec:evaluation_methodology}

To ensure robust and unbiased assessment of our parameter sweep results, we implement a controlled evaluation framework with two key components: representative test set construction and consistent subset sampling for fair model comparison.

\subsubsection{Representative Test Set Construction}
\label{sec:representative_test_set}
We construct our test set using \textbf{stratified random sampling} to maintain the same TOI-to-total light curve ratio as observed in the full sector data. This approach ensures that performance metrics remain meaningful and are not artificially inflated or deflated by over- or under-representation of confirmed TOIs in the evaluation set. The stratified sampling preserves the natural class distribution, enabling fair comparison of model performance against realistic detection scenarios.

\subsubsection{Controlled Subset for Model Comparison}
\label{sec:controlled_subset}
To enable computationally efficient and methodologically sound comparison across all 36 URF sub-variant models, we employ a fixed subset approach for both training and evaluation. A \textbf{randomly selected stratified subset of 4,000 light curves is drawn from the full SPOC 2-minute cadence sector 2} dataset and used consistently across all parameter combinations. This controlled framework offers several advantages:

\begin{itemize}
    \item \textbf{Computational efficiency:} Reduces training and evaluation time across the extensive parameter grid
    \item \textbf{Controlled comparison:} Eliminates variation due to different light curve content across experimental runs
    \item \textbf{Isolated parameter effects:} Ensures performance differences are attributable solely to synthetic set design choices rather than random sampling variation
    \item \textbf{Trend Analysis Step}: The performance of the 36 models on this small subset helps us analyze trends in their behavior in relation to the parameters of synthetic set generation and quantify those trends using the performance metrics in \ref{sec:performance_metrics} and then test the generalization of those trends on independent sectors using cross sector evaluation of the models with more (and better, in terms of representation of quantification of model performance) performance metrics described in Table \ref{tab:alpha_metrics} 
\end{itemize}

\subsubsection{TOI Importance Weighting }
\label{sec:toi_importance_weighting}
To prioritize TOIs by their detectability and astrophysical significance, we define an importance score that favors targets with stronger observational signatures:
\begin{equation}
\text{Importance Score} = \text{SNR} \times \text{Depth (ppm)} \times \text{Duration (days)}
\label{eq:importance_score}
\end{equation}

where SNR represents the signal-to-noise ratio of the planet detection, Depth (ppm) is the transit depth in parts per million, and Duration (days) is the transit duration in days. This heuristic emphasizes TOIs that are deeper, longer-duration, and higher signal-to-noise, making them both more detectable and more reliable for anomaly detection evaluation. For fair comparison across sectors, TOI importance scores are normalized to the range $[0,1]$.

This normalization enables sector-independent comparison, with higher normalized scores indicating greater relative astrophysical importance within the evaluation set. The TOI Importance AUC metric then rewards models that assign high anomaly scores to the most observationally significant targets, providing a more nuanced assessment than simple recall alone. \newline 

\subsubsection{Performance Metrics}
\label{sec:performance_metrics}

We evaluate URF model performance using three complementary metrics designed to capture both detection efficiency and astrophysical relevance:

\begin{itemize}
    \item \textbf{Anomaly Rate:} Fraction of the test set flagged as anomalous, providing insight into model selectivity
    \item \textbf{TOI Recall:} Fraction of known TOIs successfully recovered, measuring detection completeness
    \item \textbf{TOI Importance AUC:} Area under the curve when TOIs are ranked by anomaly score and weighted by astrophysical importance as described in section \ref{sec:toi_importance_weighting}
\end{itemize}

We emphasize that TOIs are \textit{never used in training or hyperparameter tuning}. They serve solely as an independent performance metric. Cross-sector tests (Section \ref{sec:cross_sector_alpha}) provide additional validation of model stability, while clustering and cosine similarity analyses (Section \ref{sec:astrophysicalrelevanceandmorphologicalclusteringofurfanomaliesfortransitdetection}) are presented only for contextual interpretation, not model selection. This ensures that our evaluation avoids circular validation and remains consistent with the unsupervised framework. \newline

\textbf{Threshold-Based Recall Analysis:} To assess how effectively each URF-4 sub-variant ranks TOIs among detected anomalies, we perform systematic threshold analysis across varying anomaly score cutoffs. For each threshold corresponding to the top $n\%$ of anomaly scores (where $n \in \{5, 10, 15, \ldots, 100\}$), we calculate:

\begin{equation}
R_n = \frac{|\{i : i \in \text{top } n\% \text{ by anomaly score and } i \in \text{TOI}\}|}{|\text{TOI test set}|}
\label{eq:toi_recall_threshold}
\end{equation}

This analysis reveals whether TOIs are preferentially concentrated at higher anomaly scores, indicating effective ranking behavior, and quantifies how recall performance varies as detection thresholds are adjusted.

It is worth noting that TOI Importance AUC (and other importance metrics derived further on) and TOI Recall serve as astronomy-contextualized analogs of traditional precision-recall metrics in machine learning performance analysis. While standard recall measures the fraction of positive class instances correctly identified, TOI Recall specifically quantifies the fraction of  confirmed/nominated transit candidates recovered by the anomaly detection framework. Similarly, TOI Importance AUC extends beyond simple area-under-curve calculations by weighting detections according to their observational significance, providing a domain-specific assessment of model ranking quality that prioritizes the most scientifically valuable targets.

\subsection{Combined Metric Analysis}
\label{sec:combinedmetricanalysis1}

To optimize synthetic light curve parameters for specific scientific objectives, we introduce a combined performance metric that balances detection completeness against astrophysical relevance:
\textbf{
\begin{align}
\text{S}_{\text{combined}} &= \alpha \cdot \text{Recall AUC}_{\text{TOI}} \notag \\
&\quad + (1 - \alpha) \cdot \text{Importance AUC}_{\text{TOI}}
\label{eq:combined_score1}
\end{align}
}

The weighting parameter $\alpha$ enables systematic control over the scientific trade-off between casting a \textbf{wider net} (high $\alpha$, emphasizing comprehensive TOI recall) versus casting a \textbf{more accurate net} (low $\alpha$, prioritizing astrophysically significant (higher values of importance metrics) targets).

For a given $\alpha$, the combined score \textbf{$\text{S}_{\text{combined}}$} is calculated for each urf-4 variant model and then ranked to select the top model. \newline

This formulation satisfies the following properties:
\begin{enumerate}
    \item \textbf{Convexity:} Since $\alpha \in [0,1] $  $S_{\text{combined}} $ lies within the convex hull of $R$ and $I$. This guarantees that the combined score is always bounded between the two metrics and cannot extrapolate beyond them i.e. $ S_{\text{combined}} \in [\min(R, I), \max(R, I)]$.
    \item \textbf{Continuity:} The score varies smoothly as $\alpha$ changes, enabling gradient-free optimization (e.g., grid search) over \( \alpha \) values.
    \item \textbf{Trade-off Interpretation:} 
    \begin{itemize}
        \item $\alpha \to 1$ emphasizes \textbf{Recall AUC}, optimizing for high completeness (i.e., casting a wide net to detect as many TOIs as possible).
        \item $\alpha \to 0$ emphasizes \textbf{Importance 
AUC}, favoring models that score astrophysically significant TOIs highly, even at the cost of total recall. A more comprehensive counterpart of this combined scoring metric has been presented in Equation \ref{eq:final_combined_scoring_metric}. \end{itemize} \end{enumerate}

In Part II of this paper, we introduce a more comprehensive version of this combined score, adjusted for a total of six performance metrics, that satisfy the same properties as listed above.

\subsubsection{Generalization Evaluation of Balanced URF Variant ($\alpha$=0.5)}
\label{sec:generalization_evaluation}

To validate the stability and generalization of our balanced URF-4 configuration for $\alpha$=0.5, we evaluate the balanced model across multiple random subsets of unseen data compiled with stratified sampling from the same sector (sector 2). In this context, by “generalization” we refer to how consistent a model’s performance is across 10 randomly sampled, stratified subsets of the same sector, that is, its stability over similar data distributions. Specifically, we compute the mean and deviation of performance metrics relative to predictions made from the trend-prediction test set defined in Section \ref{sec:controlled_subset}. \newline

\textbf{Selected Model ($\alpha = 0.5$):} \texttt{urf4\_n200\_d13\_c2\_n50ppm}

\textbf{Target Performance Metrics (predicted from trend analysis results):}
\begin{itemize}
    \item TOI Recall AUC $\approx$ 0.561
    \item TOI Importance AUC $\approx$ 0.120
\end{itemize}

\textbf{Evaluation Protocol:}
Our generalization assessment employs a rigorous random sampling approach:
\begin{enumerate}
    \item \textbf{Random Sampling:} Generate 10 disjoint/overlapping subsets of 4,000 light curves each from the full sector 2 SPOC dataset.
    \item \textbf{Anomaly Scoring:} Apply the selected URF model to compute anomaly scores for each subset.
    \item \textbf{Subset Analysis:} For each subset, compute TOI Recall AUC and TOI Importance AUC using known TOI classifications.
    \item \textbf{Statistical Aggregation:} Calculate mean and deviation of mean from predicted value across all subsets to assess stability.
\end{enumerate}

\textbf{Pass Criteria:}
A model demonstrates acceptable generalization if:
\begin{itemize}
    \item Both metrics exhibit low variance from predicted values(deviation < 10\% of mean)
    \item Mean performance remains within ±0.03–0.05 of target values
\end{itemize}

Results of this generalization has been explored in Section \ref{sec:res_generalization_alpha_p5}.

\begin{table*}
\centering
\small
\begin{tabular}{|l|l|l|}
\hline
\begin{minipage}[t]{3cm}\textbf{Metric}\end{minipage}
& \begin{minipage}[t]{4cm}\textbf{What It Measures}\end{minipage}
& \begin{minipage}[t]{6cm}\textbf{Why It Matters}\end{minipage} \\
\hline

\begin{minipage}[t]{3cm}\raggedright TOI Recall AUC\end{minipage}
& \begin{minipage}[t]{4cm}\raggedright Distribution quality of TOIs toward top anomaly scores\end{minipage}
& \begin{minipage}[t]{6cm}\raggedright Reflects overall prioritization quality across the full score spectrum\end{minipage} \\
\hline

\begin{minipage}[t]{3cm}\raggedright TOI Importance AUC\end{minipage}
& \begin{minipage}[t]{4cm}\raggedright Preservation of high-value TOIs across score thresholds\end{minipage}
& \begin{minipage}[t]{6cm}\raggedright Indicates whether scientifically valuable objects are ranked highly\end{minipage} \\
\hline

\begin{minipage}[t]{3cm}\raggedright Mean Importance (Top 20\%)\end{minipage}
& \begin{minipage}[t]{4cm}\raggedright Average importance of TOIs within top 20\% of scores\end{minipage}
& \begin{minipage}[t]{6cm}\raggedright Shows strength of early-ranked candidates --- critical for targeted follow-up\end{minipage} \\
\hline

\begin{minipage}[t]{3cm}\raggedright Binary TOI Recall\end{minipage}
& \begin{minipage}[t]{4cm}\raggedright Fraction of TOIs flagged as anomalies (score $>$ 0)\end{minipage}
& \begin{minipage}[t]{6cm}\raggedright Direct measure of detection breadth for candidate mining\end{minipage} \\
\hline

\begin{minipage}[t]{3cm}\raggedright Anomaly Rate\end{minipage}
& \begin{minipage}[t]{4cm}\raggedright Fraction of all light curves flagged as anomalous\end{minipage}
& \begin{minipage}[t]{6cm}\raggedright Reflects follow-up cost --- high rates imply more manual vetting\end{minipage} \\
\hline

\begin{minipage}[t]{3cm}\raggedright Percentile Range (Top N\%)\end{minipage}
& \begin{minipage}[t]{4cm}\raggedright Anomaly score percentile range containing top N\% of TOIs by Importance Score\end{minipage}
& \begin{minipage}[t]{6cm}\raggedright Measures how early high-value objects appear --- useful for threshold setting\end{minipage} \\
\hline

\end{tabular}
\caption{Six-metric evaluation framework for $\alpha$-dependent behavior analysis}
\label{tab:alpha_metrics}
\end{table*}

\subsubsection{Hypothesis Validation: Alpha Variants}
\label{sec:alpha_variants}

To validate the hypothesis that synthetic set design can be used to \textbf{steer model behavior toward different scientific objectives}, two additional URF-4 subvariants are evaluated over 10 sets of 4,000 light curves each  -  totaling \textbf{40,000 light curves per model} (the same subsets used to evaluate the balanced model with $\alpha$ = 0.5)  -  with configurations selected based on $\alpha$-value ranking:

\begin{itemize}
    \item \textbf{Low-$\alpha$ model} ($\alpha$ = 0.3): Emphasizes astrophysical importance
    \item \textbf{High-$\alpha$ model} ($\alpha$ = 0.9): Prioritizes detection completeness
\end{itemize}

\textbf{Goal:} Assess whether changing $\alpha$ systematically alters model behavior across multiple evaluation dimensions  -  not just recall versus importance, but also prioritization sharpness, anomaly rate, and TOI distribution in score space. We then extend this across five separate sectors to test generalization on independent populations.  \textbf{Model selection by this $\alpha$ metric is subject to a light supervisory step as referred to in Appendix \ref{sec:noteonmodelselectionstrategy}.} \newline

\textbf{Evaluation Framework: Six-Metric Assessment} \newline

URF-4 $\alpha$ variants are evaluated across six complementary metrics, each capturing different aspects of anomaly detection performance and scientific utility as described in Table \ref{tab:alpha_metrics}. Results of this validation test for alpha variants working as per design have been discussed in Section \ref{sec:interpretationofalphacomparisonacrosssixmetricssamesector}.

\subsubsection{Cross-Sector Generalization of Alpha Variants}
\label{sec:cross_sector_alpha}

We next evaluate \textbf{cross-sector generalization} to test whether these behavioral patterns (Section \ref{sec:interpretationofalphacomparisonacrosssixmetricssamesector}) persist across different TESS observational periods. This analysis addresses a critical question for practical deployment: do models selected and characterized on one sector maintain their performance characteristics when applied to entirely different stellar populations and observing conditions? \newline

\textbf{Evaluation Strategy:} Each $\alpha$ variant (0.3, 0.5, 0.9) is evaluated on the same six-metrics tracked across 5 TESS SPOC sectors S1, S2, S3, S4 and S5 to establish:
\begin{enumerate}
    \item \textbf{Performance stability:} Whether relative $\alpha$-dependent behaviors remain consistent across sectors
    \item \textbf{Best performer identification:} Which $\alpha$ value optimizes each performance metric for different stellar populations
\end{enumerate}

For each sector, 10 random stratified subsets for Sectors 1 and 2, and 4 subsets for Sectors 3, 4, and 5, sample sets of 4000 light curves each were evaluated using the three $\alpha$-variant models. Sector-level means of the six performance metrics were then computed for comparative analysis. \textbf{In total, each model variant scored exactly 128{,}000 light curves across the five sectors, resulting in an aggregate of 384{,}000 curves scored by an urf model across all $\alpha$ variants for this cross-sector generalization test.} \newline

The means of each of the six metrics have been presented in Figure \ref{fig:alpha-variants-cross-sector-test-performance-metrics-values}. The best performing alpha variant across the metrics have been presented in Table \ref{tab:crosssectoralphavariantperformance}. Results have been discussed in Section \ref{sec:res_cross_sector_gen_alpha_var}.

\subsection{Benchmarking URF Models}
\label{sec:part1modelbenchmarking}

\begin{figure}
    \centering
    \includegraphics[width=1\linewidth]{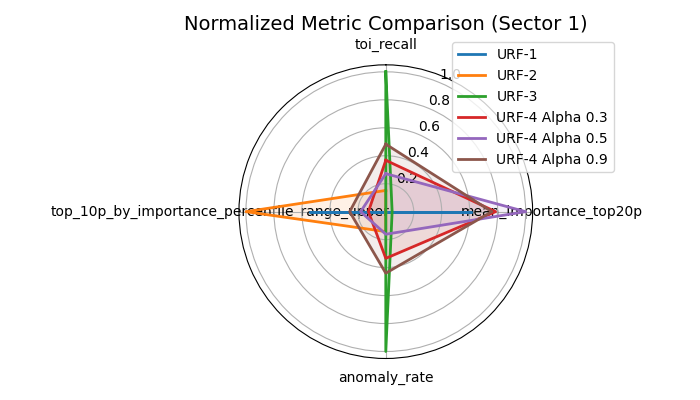}
    \caption{Normalized metric comparison across URF variants for Sector 1. URF-4 $\alpha$ models outperform fixed baselines and allow strategic tuning of performance based on discovery goals.}
    \label{fig:radar_sector1}
\end{figure}

To evaluate the impact of controllable synthetic design, we benchmark the performance of our URF-4 $\alpha$-variants against previously established baselines: URF-1 (uniform noise), URF-2 (fixed hyperparameters from MG23), and URF-3 (TOI-based synthetic set). Figure~\ref{fig:radar_sector1} shows a radar plot for Sector 1 comparing URF variants across normalized metrics including TOI recall, anomaly rate, importance-weighted metrics, and variability in top-importance scores. Ideal models minimize \texttt{top\_10p\_by\_importance\_percentile\_range\_upper}, strike a trade-off between \texttt{toi\_recall} and \texttt{anomaly\_rate}, and maximize \texttt{mean\_importance\_top20p}. As evident by the radar plot, URF-4 occupies the goldilocks zone compared to the performance of URF 1-3.\newline

\subsection{Part 2 - Astrophysical Relevance and Morphological Clustering of URF Anomalies For Transit Detection}
\label{sec:astrophysicalrelevanceandmorphologicalclusteringofurfanomaliesfortransitdetection}

Having established optimal URF architectures and validated behavioral steering mechanisms, in this section, we shift focus from model selection and parameter optimization to the astrophysical validation and interpretability of anomalies detected by the CLARA framework. We now seek to answer a fundamental question: \emph{Do the anomalies identified by our optimized models correspond to genuine astrophysical phenomena, or are they dominated by noise and instrumental artifacts? } \newline

To address this first, we perform a feature-to-metric mapping over a finer mesh of synthetic set feature values as follows, systematically varying synthetic light curve parameters (such as noise, and duration, as they were the most important feature according to the Random Forest Regressor) to pick 0.3 and 0.9 $\alpha$ variant urf-4 models as described in the previous section.
 
\begin{itemize}
\item \textbf{Noise levels (ppm):} [25, 50, 75, 100, 125, 150, 175, 200, 225, 250, 275, 300, 400, 600, 800]
    \item \textbf{Duration (days \ref{sec:noteonduration}):} [0.73, 1.65, 3.0, 3.58, 5.1, 6.0, 7.0, 8.3, 10.0, 12.0, 13.0, 15.67, 17.0, 20.0, 21.0, 21.5, 22.0, 22.5, 23.0, 23.25, 23.5, 23.75, 24.0, 24.25, 24.5, 24.75, 25.0, 25.25, 25.5, 26.0, 27.0]
    \item \textbf{Fixed parameters:} n\_curves = 300, cadence\_minutes = 2
\end{itemize}

Second, we rigorously validate the astrophysical significance of detected anomalies. By correlating anomaly scores with known stellar and planetary properties, and employing techniques such as Normalized Weighted Root Sum of Squares (N-WRSS) scoring and t-SNE clustering \citep{vanDerMaaten2008}, we map the distribution of high-scoring anomalies in both feature and astrophysical parameter space. Furthermore, we leverage external catalogs such as SIMBAD \citep{Wenger2000} to perform morphological classification, quantitatively demonstrating that our models are sensitive to real transit-like events and astrophysically meaningful outliers. \newline

Together, these analyses establish a robust framework for interpreting and validating unsupervised anomaly detection in large-scale time-series surveys, providing both scientific insight and a blueprint for extending CLARA to new domains and datasets.

\subsection{A Finer Mapping of Features to Performance Metrics}
\label{sec:finermappingsec2a}

These two parameters — \textbf{noise (ppm)} and \textbf{transit duration} — were selected for finer-grained parameter grid variation tests (over the values mentioned in Section \ref{sec:astrophysicalrelevanceandmorphologicalclusteringofurfanomaliesfortransitdetection}, based on feature importance analysis using the random forest regressor from earlier URF-4 performance modeling (Figures~\ref{fig:feature_importance_comparison},~\ref{fig:TOIImportanceAUCvsSynthParamFeatureImportance}).  The grid was intentionally densified in the ranges of \textbf{duration $\sim$23–25.5 days} and \textbf{noise $\sim$150–300 ppm}, where prior experiments indicated sharp performance transitions, enabling finer resolution of model behavior in these critical regions. \newline

\begin{figure}
    \centering
    \includegraphics[width=1\linewidth]{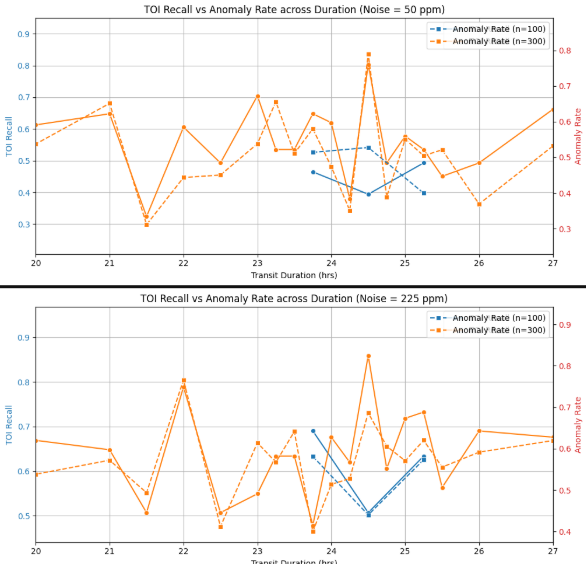}
    \caption{Recall vs Anomaly Rate across Duration values for Noise = 50, 225. Yellow line is  for n\_curves=300, blue for n\_curves=100}
    \label{fig:durationvariationfinermappingzoomed}
\end{figure}

Noise (ppm) values \textbf{50e-6 and 225e-6 turned out most dominant in terms of importance}. The 10P$_{importance}$ Upper Limit values for the most dominant noise values were 1.70 and 34.22 respectively. Fixing these values for noise, we varied duration for which the range from 20 days to 26 days showed predictable variability for both noise values. (Noise (ppm) = 100 was also tested for the three duration values for $\alpha$=.3, $\alpha$=.5, and $\alpha$=.9 models) as shown in Figure \ref{fig:durationvariationfinermapping} and \ref{fig:durationvariationfinermappingzoomed} (figure \ref{fig:durationvariationfinermapping} scaled up to 20 days to 27 days).
Final alpha variant model selections were done using this revised combined scoring metric:
\begin{align} \label{eq:final_combined_scoring_metric}
S_{\text{combined}}(\alpha) &= w_{\text{recall}} \cdot R_{\text{norm}} + w_{\text{anomaly}}^{-1} \cdot A_{\text{inv}} + w_{\delta} \nonumber \\
&\quad \cdot \Delta_{\text{norm}} + w_{\text{importance}} \cdot I_{\text{norm}} + w_{\text{range}} \cdot P_{\text{inv}}
\end{align}
where the weight assignments follow an $\alpha$-controlled group structure: \newline

Detection Group (weight = $\alpha$):
\begin{align}
w_{\text{recall}} &= 0.5 \alpha \\
w_{\text{anomaly}}^{-1} &= 0.2 \alpha \\
w_{\delta} &= 0.3 \alpha
\end{align}

Interpretability Group (weight = $1 - \alpha$):
\begin{align}
w_{\text{importance}} &= 0.5 (1-\alpha) \\
w_{\text{range}} &= 0.5 (1-\alpha)
\end{align}

and the normalized/inverted metrics are defined as: 
\begin{align}
R_{\text{norm}} &= \text{MinMaxScaler}(\text{TOI Recall}) \\
A_{\text{inv}} &= 1 - \text{MinMaxScaler}(\text{Anomaly Rate}) \\
\Delta_{\text{norm}} &= \text{MinMaxScaler}(\text{Recall-Anomaly Delta}) \\
I_{\text{norm}} &= \text{MinMaxScaler}(\text{Mean Importance Top 20\%}) \\
P_{\text{inv}} &= 1 - \text{MinMaxScaler}(\text{Top 10\% Range Upper})
\end{align}
This revised scoring function retains convexity: all five component metrics are normalized to $[0, 1]$ and weighted by non-negative coefficients summing to 1, yielding a bounded, interpretable score. The weights are structured such that the total contribution from the Detection group is controlled by $\alpha$, and the Interpretability group by $(1 - \alpha)$, preserving the convex combination property and ensuring no extrapolation beyond the empirical performance ranges.

The revised scoring metrics is justified in terms of maximizing the weighted sum of five components as follows: \newline

\textbf{Detection Group ($\alpha$ weight):}

\begin{enumerate}
    \item TOI Recall (0.5$\alpha$) - maximize detection performance
    \item Inverted Anomaly Rate (0.2$\alpha$) - minimize false positives
    \item Recall-Anomaly Delta (0.3$\alpha$) - maximize gap between recall and anomaly rate

\end{enumerate}

\textbf{Interpretability Group ($1-\alpha$ weight):}

\begin{enumerate}
    \item Feature Importance (0.5($1-\alpha$)) - maximize clarity of top 20\% features
    \item Inverted Range Upper (0.5($1-\alpha$)) - minimize prediction instability in top 10\%
\end{enumerate}

A formal derivation of the combined scoring metric and its $\alpha$-controlled weighting scheme remains an open direction for future work as discussed under the banner of \textbf{Theoretical Framework Development} section of future works discussion (\ref{sec:future_work}). \newline

Following are the urf-4 alpha variant models that were picked for the three alpha values: 
\begin{center}
\begin{itemize}
    \item $\alpha_{0.3}$ - urf4\_model\_variant\_n300\_d23.75\_c2\_n50ppm
    \begin{itemize}
        \item Predicted: TOI Recall = 0.64; Anomaly Rate = 0.58; Mean Importance Top 20P = 0.42, Top 10P$_{importance}$ Upper Limit = 2.5
    \end{itemize}
    \item $\alpha_{0.5}$ - urf4\_model\_variant\_n300\_d25.25\_c2\_n225ppm 
    \begin{itemize}
        \item Predicted: TOI Recall = 0.73; Anomaly Rate = 0.62; Mean Importance Top 20P = 0.45, Top 10P$_{importance}$ Upper Limit = 19.7
    \end{itemize}
    \item $\alpha_{0.9}$ - urf4\_model\_variant\_n300\_d24.5\_c2\_n225ppm 
    \begin{itemize}
        \item Predicted: TOI Recall = 0.85; Anomaly Rate = 0.68; Mean Importance Top 20P = 0.37, Top 10P$_{importance}$ Upper Limit = 81.9
    \end{itemize}
\end{itemize}
\end{center}

\subsection{Weighted Root Sum Score: A revision of URF anomaly scores}
\label{sec:wrss}

Ensemble fusion provides a natural way to balance the complementary behaviors of URF models with different $\alpha$ values, combining conservative precision and aggressive recall. In particular, the $\alpha=0.3$ and $\alpha=0.9$ models span a broad regime, from conservative anomaly flagging to strong recall, making their intersection an informative testbed for robust score combination. \newline

As observed previously, the low $\alpha$ model is conservative over flagging TOIs and does better at scoring important curves higher. Meanwhile, the high $\alpha$ model is significantly more aggressive in flagging curves and casts a wider net with higher TOI recall. To allow ourselves hybrid control over aggressiveness and conservativeness, we needed a combined scoring function for the intersection set of the anomalies (non zero anomaly score curves) flagged by these two models. Thus, we pick the weighted root sum score (WRSS):
\begin{equation}
    s_w = w_3\, \sqrt{s_{3}'} + w_9\, \sqrt{s_{9}'}
\end{equation}

where $s_{3}'$ and $s_{9}'$ denote the raw anomaly scores from models $\alpha = 0.3$ and $\alpha = 0.9$ respectively, and $s_w$ is the weighted score computed from their square roots. The weights $w_3$ and $w_9$ satisfy: 
\begin{equation}
    w_3 + w_9 = 1
\end{equation} To give higher preference to better scoring, we set the weights as $w_3  = 0.8$ and $w_9 = 0.2$. \newline

For a generalization of the N-WRSS, let there be $M$ anomaly detection models indexed by $i \in \{1, 2, \dots, M\}$, each producing a raw anomaly score $s_i'$ for a given light curve. We define the \textbf{generalized weighted root sum score} (WRSS) for the intersection set of anomalies flagged by these models as:

\begin{equation}
    s_w = \sum_{i=1}^{M} w_i \, \sqrt{s_i'}
\end{equation}

where $w_i \in [0,1]$ are the weights assigned to each model such that:
\begin{equation}
    \sum_{i=1}^{M} w_i = 1
\end{equation}

The choice of $w_i$ allows control over the trade-off between aggressive and conservative anomaly flagging by emphasizing models with desired properties.

For standardized comparison across runs on different test sets, we define the \textbf{generalized normalized WRSS} (N-WRSS) as:
\begin{equation}
    \hat{s}_w = \frac{s_w - \min(s_w)}{\max(s_w) - \min(s_w)}
\end{equation}
where $\min(s_w)$ and $\max(s_w)$ are computed over all curves in the considered set.

\subsubsection{Why compute N-WRSS for $\alpha_{0.3}$ - $\alpha_{0.9}$ intersection sets?}

The \textbf{normalized weighted root sum score (N-WRSS $\hat{s}_w$)} provides a \textbf{robust ensemble score fusion}, capturing the degree of agreement between different URF models ($\alpha_{0.3}$ and $\alpha_{0.9}$ in this case). By computing $\hat{s}_w$ over the intersection set — i.e., light curves flagged by both models — we:

\begin{itemize}
    \item \textbf{Quantify inter-model consistency}: A low $\hat{s}_w$ implies the two models assign similar anomaly scores (after normalization), reinforcing confidence in the flagged anomalies.
    \item \textbf{Mitigate single-model dominance}: Linear combinations of normalized scores can be dominated by extreme values from a single model, particularly when models differ in sensitivity or training noise. Applying a \textit{monotone concave transformation} (e.g., square root) dampens extreme scores, preserving rank order while ensuring the N-WRSS reflects \textit{consensus} rather than outlier-driven influence. \textbf{Mathematical intuition:} For a monotone concave transformation $f$, we have \begin{equation}
        f(x)-f(y) \le f'(y)(x-y),\end{equation} with \begin{equation}f'(y)<1
    \end{equation} This shows that large differences between extreme scores are compressed, preserving rank order while reducing outlier influence.

    \item \textbf{Support threshold stability}: Dampening extremes enables more robust fine-tuning of N-WRSS thresholds across different datasets, improving reproducibility of anomaly detection.
\end{itemize}

The empirical benefits of this formulation, in contrast to a simple linear weighted sum score, are demonstrated through Monte Carlo injection-recovery validation in Section \ref{sec:mc_eval_metrics} and Figure \ref{fig:linear_vs_nwrss}, where N-WRSS is shown to sustain higher recall and more stable precision-recall tradeoffs across thresholds. \newline

Thus, $\hat{s}_w$ on the $\alpha_{0.3}$ - $\alpha_{0.9}$ intersection forms a critical bridge for \textbf{robust ensemble scoring}, providing a stable consensus measure before downstream similarity analysis or labeling. A fuller theoritical treatment of the concavity framework and its implication for ensemble justification is deferred to Section \ref{sec:future_work}.

\subsection{Astrometric and Stellar Parameter Augmentation}
\label{sec:astro_features}

To enable physical interpretation of anomalies detected by URF-4, we enrich high-scoring candidates with astrophysical and astrometric metadata. Specifically, we extract relevant features from the \textit{TESS Input Catalog} (TIC) and \textit{Gaia DR3} for all candidates which exceeded a chosen anomaly threshold (in our case $N-WRSS \ge 0.35$).

We begin by parsing TIC IDs from each candidate's filename. These IDs are used to query the MAST TIC catalog to retrieve:

\begin{itemize}
    \item \textbf{Astrometry:} parallax (\texttt{plx}), proper motion in right ascension and declination (\texttt{pmra}, \texttt{pmdec})
    \item \textbf{Stellar parameters:} effective temperature (\texttt{Teff}), radius (\texttt{rad}), mass (\texttt{mass}), luminosity (\texttt{lum}), and surface gravity (\texttt{logg})
\end{itemize}

To further characterize stellar kinematics, we compute the tangential velocity \( v_{\text{tan}} \), defined as:
$$
v_{\text{tan}} = 4.74 \times \frac{\sqrt{\mu_{\text{RA}}^2 + \mu_{\text{DEC}}^2}}{\pi}
\label{eq:vtan}
$$
where \( \mu_{\text{RA}}, \mu_{\text{DEC}} \) are the proper motions in milliarcseconds per year, and \( \pi \) is the parallax in milliarcseconds.

To assess the reliability of Gaia astrometric solutions, we query the Gaia DR3 catalog for the Renormalized Unit Weight Error (RUWE) of each source, a proxy for fit quality and potential binarity or extended structure.

The resulting dataset comprises both anomaly metadata and astrophysical properties, enabling cluster-level interpretation. The full set of appended features includes:

\begin{center}
\texttt{plx}, \texttt{pmra}, \texttt{pmdec}, \texttt{Teff}, \texttt{rad}, \texttt{mass}, \texttt{lum}, \texttt{logg}, \( v_{\text{tan}} \), \texttt{ruwe}
\end{center}

Refer to Appendix \ref{sec:astrostellarfeaturedesc} for a descriptive list of these features.

\subsection{Evaluating Astrometric Feature Discriminability}
\label{sec:tsne_silhouette}

To assess which combinations of astrometric and stellar parameters most effectively distinguish high-anomaly candidates, we perform a t-SNE embedding on Sector 1 URF-4 $\alpha_{0.3} \cap \alpha_{0.9}$ set.

We generate all combinations of two or more features from the enriched feature set (Section~\ref{sec:astro_features}), and for each subset:

\begin{enumerate}
    \item Apply standard scaling to the selected features after removing NaN entries.
    \item Use t-SNE to project the high-dimensional feature space into two dimensions.
    \item Label each point as anomalous or not, based on a binary threshold (0.6 (arbitrary) for our case) on the URF-derived N-WRSS.
    \item Compute the silhouette score for the binary labels in the embedded space.
\end{enumerate}

This process gives an intuitive insight into discriminative physical features for downstream interpretation, linking anomaly detection outcomes to astrophysical structure. \newline

This method provides an initial quantitative basis for linking astrometric and stellar parameters to anomaly clustering behavior. Future work (\ref{sec:future_work}) will expand this into a formal statistical validation framework, including comparison with established methods (e.g. \citet{Crake2023}; \citet{Melton2024a}) and deeper analysis of why certain features like \textit{T}$_\mathrm{eff}$, RUWE, and $v_\mathrm{tan}$ exhibit strong correlations with high N-WRSS anomalies, thereby enabling more grounded astrophysical interpretation of detected outliers. \newline

The results of stellar parameter based t-SNE clustering with N-WRSS overlay showing the correlation between URF scoring and astrophysical features have been demonstrated and discussed in Section \ref{sec:stellar_tsne_results}.

\subsection{SIMBAD Weak Supervision and Morphological Similarity Matching}
\label{sec:COSINEMATCHINGINTROSEC}

To bridge anomaly detection with astrophysical classification, we introduce a weakly supervised labeling approach using the SIMBAD object taxonomy. A subset of high-anomaly light curves (TOIs from Sector 2 - specific choice of sector in this case is arbitrary - for which Simbad Object Labels could be found) is cross-matched with MAST astronomical coordinates to retrieve SIMBAD labels such as \texttt{Pl}, \texttt{EB*}, \texttt{V*}, and others. This offers valuable physical priors without introducing strong biases. These labeled instances are used to construct a Sector 2 reference set of TOIs with known SIMBAD classifications, enabling downstream morphological matching and interpretation. Cosine similarity in the learned feature space propagates these labels to the larger unlabeled set, in line with graph-based semi-supervised learning paradigms (\citet{vanEngelen2020}). \newline

We extract feature vectors from these labeled TOIs using Lomb-Scargle periodograms and flux values of the curves as described in \ref{sec:curve_feature}, apply PCA \citet{Hotelling1933} for dimensionality reduction, and save the compressed representations. These features form a labeled reference basis for morphological similarity-based classification. \citet{Seo_2023} \newline

Subsequently, we compute pairwise cosine similarity between urf-4 $\alpha_{0.3} \cap \alpha_{0.9}$ set candidates and this labeled reference set. Each anomaly is assigned the closest-matching SIMBAD label group based on morphological similarity of the light curve. Following label groups have been defined for this purpose:

\begin{verbatim}
    label_groups = {
    "planet_like": ["Pl", "Pl?", "BD*", "s*b"],
    "binary_star": ["SB*", "**", "EB*", "LM*"],
    "stellar": ["*", "PM*", "Er*"],
    "outlier": ["outlier", "err"]
}
\end{verbatim}

The \texttt{planet\_like} label group comprises synthetic light curves designed to exhibit transit-like morphology i.e., localized, symmetric flux dips with flat out-of-transit baselines. These shapes mimic the expected photometric signature of transits and are intended to guide anomaly detection models toward identifying signals with similar temporal structures. By concentrating on this subset, we enable both training and evaluation phases to focus on transit-relevant anomalies, filtering out unrelated variability patterns such as flares, rotational modulation, or systematics.

This step enables astrophysically interpretable annotation of previously unlabeled anomalies, allowing further insight into their likely nature and group clustering behavior. The methodology supports scalable, domain-aware classification while maintaining the unsupervised integrity of the original detection pipeline.

\subsubsection{Weak Label Acquisition via SIMBAD Cross-Matching}
\label{sec:weaklabelacquisition}

Since SIMBAD has scarce labeling when compared to the sheer number of tess curves, for a targeted and valuable labeled feature set, we chose to query the catalog for Object Type of TOIs from a specific sector. For the TIC identifier of each TOI, we queried a radius of 5 arcseconds around the right-ascension and declination for the TIC obtained from the MAST catalog. Table \ref{tab:simbad_label_freq} lists the frequencies for each label for our reference feature set.

\subsubsection{Morphological Classification via Cosine Similarity}
\label{sec:cosinesimilaritymatching}

With the Sector 2 TOIs now labeled and represented in a reduced-dimensional feature space (refer to Table \ref{tab:simbad_label_freq} for similarity reference dataset label occurence frequencies), we proceed to classify unlabeled anomalies from Sectors 1 to 5 using morphological similarity. For each anomaly in the urf-4 $\alpha_{0.3} \cap \alpha_{0.9}$ set, we extract Lomb-Scargle and flux-based features, project them using the PCA model trained on the labeled reference set, and compute cosine similarity with all reference vectors. \newline

A helper function \texttt{match\_fits\_by\_cosine\_similarity} performs the core matching step by:
\begin{enumerate}
    \item Extracting feature vectors from the input light curve using the same pipeline used for the labeled TOIs as discussed in Section \ref{sec:curve_feature}.
    \item Projecting the feature vector into the same PCA space as the reference set.
    \item Computing cosine similarity between the projected anomaly and all labeled reference TOIs.
    \item Returning the highest similarity match above a threshold, or marking the sample as an ``outlier'' if no match exceeds the threshold. \newline
\end{enumerate}

To scale this process, we implement a batch version \texttt{run\_cosine\_matching\_batch} that operates over the full anomaly candidate set using parallel workers. We apply a minimum similarity threshold (min\_sim\_threshold=0.6) to filter weak or ambiguous matches. The resulting matches are enriched with SIMBAD labels and their corresponding high-level label groups (as defined in \texttt{label\_groups}) and merged back into the anomaly dataframe. \newline

Results of cosine similarity based curve labeling have been discussed in Section \ref{sec:cosinesimilaritymatchingresults}, demonstrating, among other things, high confidence morphological similarity based classification of high N-WRSS curves into one of the labels. 

\subsection{Clustering and Morphological Filtering Methods}
\label{sec:clusteringandmorphologicalfilteringmethods}

To identify transit-like morphologies, we selected the \texttt{planet\_like} classified group of curves from TESS Sectors 1 through 5. We applied a threshold on the normalized weighted root sum square (N-WRSS) anomaly score, retaining only candidates with N-WRSS $\geq 0.35$ (arbitrary threshold based on figure \ref{fig:cosine_all} and Section \ref{sec:init_inj_rec_results}), resulting in a filtered set of 620 planet-like light curves for morphological clustering. \newline

\begin{figure*}
    \centering
    \includegraphics[width=1\linewidth]{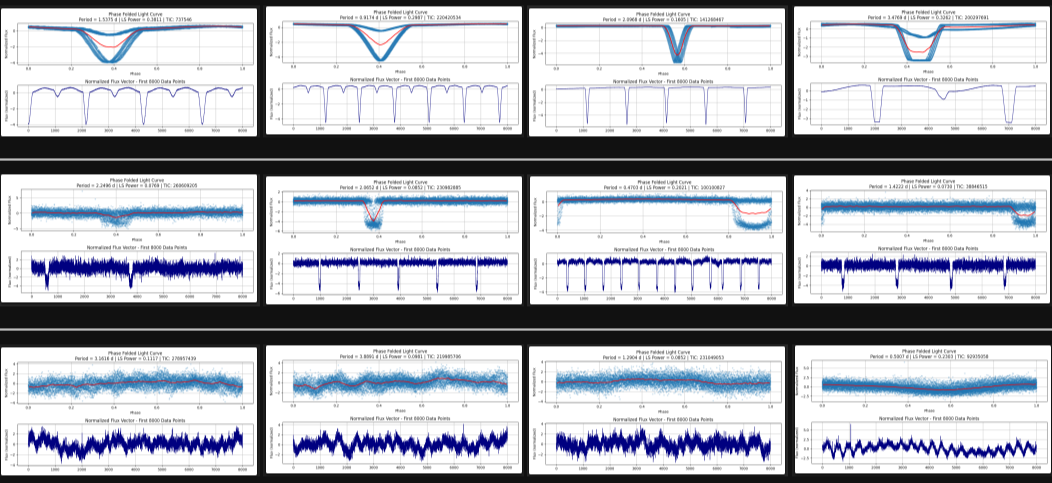}
    \caption{3 distinct morphological classes (each row represents a class) that populate the entirety of 0th cluster from DBSCAN clustering of DTW matrix}
    \label{fig:3morphclassesdtwdbscancluster0}
\end{figure*}

As a pre-filtering mechanism for clustering, we employed DBSCAN using a distance matrix derived from Dynamic Time Warping (DTW) \citet{salvador2007toward}, a technique well-suited for comparing time-series data with temporal misalignments or non-uniform sampling - conditions typical of TESS light curves. We clustered to have one large cluster with three types of morphologies as presented in Figure \ref{fig:3morphclassesdtwdbscancluster0}. \newline

To visualize the results, all curves were phase-folded using the most significant period from their Lomb-Scargle periodograms. A minimum power threshold was enforced to filter out curves with weak or unreliable periodicity. \newline

We then performed enhanced morphological feature extraction on the remaining subset, replacing the feature set described in Section~\ref{sec:curve_feature} to a refined 10-dimensional representation designed to capture morphological characteristics more effectively. \newline

Multiple clustering algorithms were evaluated on this refined feature space, and their performance was assessed both quantitatively (via TOI membership tracking) and qualitatively (via visual inspection of folded curves within each cluster). The most effective clustering approach was selected based on its ability to group plausible transit-like signals and isolate visually coherent morphology types. \newline

\subsubsection{Feature of a Curve - II}
\label{sec:curve_feature_2}

For improved clustering and morphological separation of the visually identified classes in cluster 0 (from DBSCAN applied on DTW distances), we derived a 10-dimensional feature vector for each candidate curve. This initial 10-dimensional representation serves as a baseline for morphological analysis and transit-like anomaly clustering. Future work (see Section~\ref{sec:future_work} under \textbf{Advanced Morphological Analysis}) will extend this by exploring more expressive feature combinations based on these and more. These 10 features are as follows:

\begin{enumerate}
    \item \textbf{Transit Depth} is defined as the difference between the minimum and median of the binned folded flux:

    \begin{equation}
    d = F_{\text{min}} - F_{\text{median}}
    \end{equation}

    where:
    \begin{align*}
    F_{\text{min}} &= \min_i F_i \\
    F_{\text{median}} &= \text{median}_i(F_i)
    \end{align*}

    and $F_i$ denotes the binned phase-folded flux.\newline

    \textit{Simpler explanation:} How deep the transit dip is compared to the average light level.\newline

    \item \textbf{Transit Width}: Given the folded light curve phase values $\{\phi_i\}$ and corresponding binned flux $\{F_i\}$:

    \begin{align}
    F_{\text{min}} &= \min_i F_i \\
    F_{\text{max}} &= \text{median}_i(F_i) \\
    F_{\text{half}} &= \frac{F_{\text{min}} + F_{\text{max}}}{2}
    \end{align}

    Let $\mathcal{I} = \{i \mid F_i < F_{\text{half}}\}$ be the set of indices below half-depth. Then:

    \begin{equation}
    w = 
    \begin{cases}
    \phi_{\max(\mathcal{I})} - \phi_{\min(\mathcal{I})}, & \text{if } |\mathcal{I}| \geq 2 \\
    0, & \text{otherwise}
    \end{cases}
    \end{equation}

    \textit{Simpler explanation:} The width of the transit in phase space, indicating how long the dip lasts. \newline

    \item \textbf{Baseline Standard Deviation} is computed from the top 20\% of folded flux values (assumed out-of-transit):

    \begin{align}
    \mathcal{B} &= \left\{ F_i \mid F_i > P_{80}(F) \right\} \\
    \sigma_{\text{baseline}} &= \text{std}(\mathcal{B})
    \end{align}

    \textit{Simpler explanation:} Measures the flux variation outside the transit, acting as a noise proxy. \newline

    \item \textbf{Asymmetry Index} quantifies flux asymmetry around the center of the phase curve:

    \begin{align}
    N &= \text{length of } F \\
    M &= \left\lfloor \frac{N}{2} \right\rfloor \\
    a &= \frac{1}{M} \sum_{i=1}^{M} \left| F_i - F_{N - i + 1} \right|
    \end{align}

    \textit{Simpler explanation:} Indicates whether the transit shape is symmetric or skewed. \newline

    \item \textbf{Sharpness} is the discrete curvature at the flux minimum after smoothing:

    \begin{align}
    \tilde{F} &= \text{GaussianSmooth}(F, \sigma = 3) \\
    i_{\min} &= \arg\min_i \tilde{F}_i \\
    s &= \begin{cases}
    \tilde{F}_{i_{\min} - 1} - 2\tilde{F}_{i_{\min}} \\
    \quad + \tilde{F}_{i_{\min} + 1}, & \text{if } 2 < i_{\min} < N - 3 \\
    0, & \text{otherwise}
    \end{cases}
\end{align}

    \textit{Simpler explanation:} Measures how "pointed" or sharp the dip is at its lowest point. \newline

    \item \textbf{Autocorrelation Strength} is the highest non-zero autocorrelation peak:

    \begin{align}
    \bar{F} &= \frac{F - \mu}{\sigma} \\
    \text{ACF}_k &= \sum_{i=1}^{N-k} \bar{F}_i \cdot \bar{F}_{i+k} \\
    \text{ACF}_0 &= 0 \\
    a &= \max \left\{ \text{ACF}_k \mid k > 0 \text{ and } \right. \\
    &\quad \left. \text{ACF}_k \text{ is a local maximum} \right\}
\end{align}

    \textit{Simpler explanation:} Quantifies repeating structure or periodicity in the shape of the curve. \newline

    \item \textbf{Transit Count} is the number of strong peaks in the Lomb-Scargle periodogram:

    \begin{align}
    F_{\text{norm}} &= \frac{F - \text{median}(F)}{\text{std}(F)} \\
    \text{LS}(\omega) &= \text{LombScargle}(t, F_{\text{norm}}) \\
    P(\omega) &= \text{LS.power}(\omega) \\
    \mathcal{T} &= \left\{ \omega_i \mid P(\omega_i) \text{ is a peak} \right. \\
    &\quad \left. \text{with prominence } > \delta \right\} \\
    \text{Transit Count} &= |\mathcal{T}|
\end{align}

    \textit{Simpler explanation:} Counts how many strong periodic signals are present. \newline \newline We use the Box Least Squares (BLS \citet{2002A&A...391..369K}) algorithm to identify periodic, box-shaped dips in normalized light curves—an effective method for detecting transiting exoplanets. BLS iteratively searches over trial periods to find the one that maximizes a transit signal detection statistic, fitting a simple box model to approximate the flux drop caused by a planet transit. From the best-fit period \( P^\ast \), we derive the last three of the ten interpretable metrics that characterize the transit-like signal: \newline

    \item \textbf{BLS Transit Depth} is the best-fit transit depth from a box-shaped model: 

    \begin{equation}
    d_{\text{BLS}} = \text{depth}[P^\ast]
    \end{equation}

    where \( P^\ast = \arg\max_P \text{BLS}_{\text{power}}(P) \). \newline

    \textit{Simpler explanation:} Best-estimate of how deep the box-like transit is. \newline

    \item \textbf{BLS Transit Duration} is the duration of the best-fit box-shaped transit:

    \begin{equation}
    T_{\text{dur}} = \text{duration}[P^\ast]
    \end{equation}

    \textit{Simpler explanation:} How long the transit lasts in time, according to the box model. \newline

    \item \textbf{BLS Transit Signal-to-Noise Ratio (SNR)}:

    \begin{equation}
    \text{SNR}_{\text{BLS}} = \frac{d_{\text{BLS}}}{\sigma_F}
    \end{equation}

    where \( \sigma_F = \text{std}(F_{\text{norm}}) \). \newline

    \textit{Simpler explanation:} Strength of the detected box transit signal relative to background noise. \newline
\end{enumerate}

The results and interpretation of clustering based on this new feature set has been demonstrated and discussed in Section \ref{sec:clustering_results_interp}, showing how probabilistic clustering on these method yields optimal results, as also justified empirically in Section \ref{sec:mc_eval_metrics}.

\subsection{Monte Carlo Injection-Recovery Validation}
\label{sec:monte_carlo}

A central challenge in evaluating unsupervised detection frameworks is the absence of ground-truth signals in the majority of survey light curves. Although cross-matching with TOIs provides partial validation, it is incomplete and biased toward previously identified signals. Injection-recovery testing is an established approach (e.g. \citet{2013ApJS..207...35C}, \citet{Christiansen_2020}) for quantifying the completeness and reliability of exoplanet detection pipelines. To obtain an absolute measure of sensitivity and to calibrate threshold choices, we performed Monte Carlo injection-recovery experiments using synthetic transits embedded in real TESS light curves. This approach allows us to:

\begin{enumerate}
    \item Quantify recall and vetting load under controlled conditions.
    \item Empirically justify use of NWRSS for quantification of model ensemble consensus.
    \item Empirically justify probabilistic methods like GMM for morphological clustering.
    \item analyze the false positives that survive URF selection at various thresholds.
\end{enumerate}

\subsubsection{Experimental Setup}
\label{sec:mc_setup}

We constructed a pool of 200 synthetic light curves containing transit signatures spanning a range of depths, noise levels, and orbital periods. For each run, 20 curves were randomly selected from this pool and injected into real TESS light curves. In parallel, 1,980 real SPOC light curves were sampled per run, drawn from Sectors 6–10 and 20–24 and 30, yielding a total of 2,000 curves per run. Random sampling and URF scoring were repeated 80 times to ensure statistical robustness. Both the $\alpha=0.3$ and $\alpha=0.9$ variants were applied to each run, enabling comparison of recovery performance and false-positive behavior across distinct scoring regimes. The full list of the 10 synthetic transit configurations used for injection, including depth and noise ranges, is provided in Table \ref{tab:mc_configs}. For a single Monte Carlo run, a single random configuration was picked from the ten and then 20 curves of that particular configuration were injected into the test set. A detailed evaluation of recovery performance including confidence intervals across 80 Monte Carlo trials is presented in Sections \ref{sec:init_inj_rec_results} and \ref{sec:mc_eval_metrics}.

\section{Results}

\subsection{Correlating Synthetic Parameters to Performance Metrics}
\label{sec:syntheticparamtometriccorr1}
\label{sec:syntheticparamtometriccorr1res1}

Here, we present the systematic trends we observed for these 36 urf-4 sub-variant models. Figures \ref{fig:synthparamsvsrecallanomalyrate1}, and \ref{fig:synthfeaturesvstoiimportanceauc} present the correlation between individual synthetic parameters and the performance metrics TOI Recall, Anomaly Rate and TOI Importance AUC (\ref{sec:noteontoiimpauc}). \newline

To quantify the relative influence of synthetic training parameters on URF-4 performance, we employed Random Forest regression to model the relationship between our four key synthetic parameters (\texttt{n\_transits}, \texttt{duration\_hr} (\ref{sec:noteonduration} \footnote{During the course of this research, the duration of transit-like events is consistently measured in \textbf{days} across all computations, analyses, and visualizations.}), \texttt{cadence\_min}, \texttt{noise\_ppm}) and three primary evaluation metrics (TOI Recall, Anomaly Rate and TOI Importance AUC). This approach provides interpretable feature importance scores that reveal which synthetic design choices most strongly influence model behavior.

\begin{figure*}
\centering

    \includegraphics[width=1\linewidth]{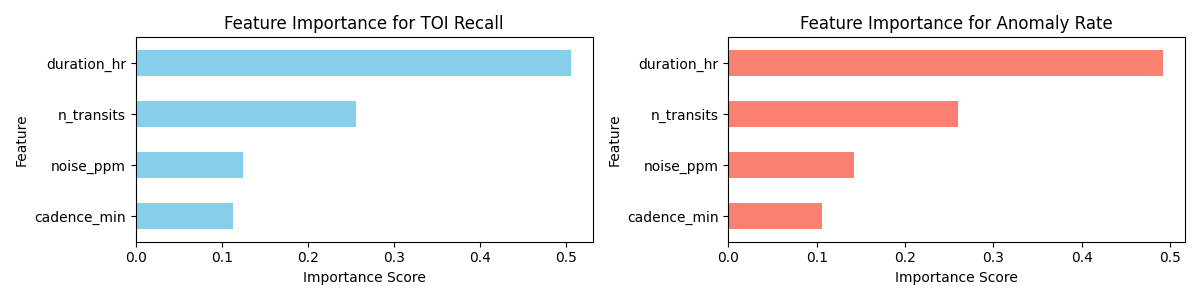}
\caption{Feature importance comparison for URF-4 synthetic training parameters. Left panel shows importance scores for TOI Recall prediction, right panel for Anomaly Rate prediction. Transit duration (days \ref{sec:noteonduration}) emerges as the dominant factor for both metrics, followed by transit multiplicity. Random Forest regressors trained on 36-variant parameter sweep results with 42-seed random state for reproducibility.}
\label{fig:feature_importance_comparison}
\end{figure*}

\begin{figure*}
    \centering
    \includegraphics[width=1\linewidth]{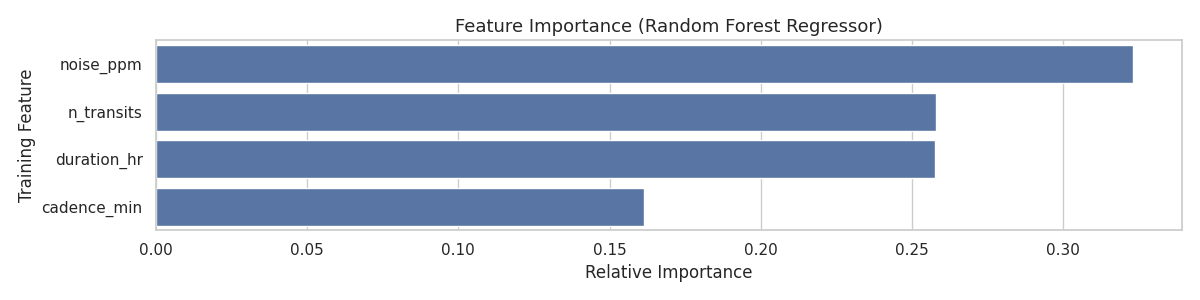}
    \caption{Feature importance comparison for URF-4 synthetic training parameters. This plots TOI Importance AUC feature importance against the features.}
    \label{fig:TOIImportanceAUCvsSynthParamFeatureImportance}
\end{figure*}

The Random Forest regression analysis, used to quantify how synthetic training parameters influenced URF-4 performance (Figure~\ref{fig:feature_importance_comparison}, \ref{fig:TOIImportanceAUCvsSynthParamFeatureImportance}), directly informed the finer-grained experiments presented in Section~\ref{sec:finermappingsec2a}.

\begin{figure*}
    \centering
    \includegraphics[width=1\linewidth]{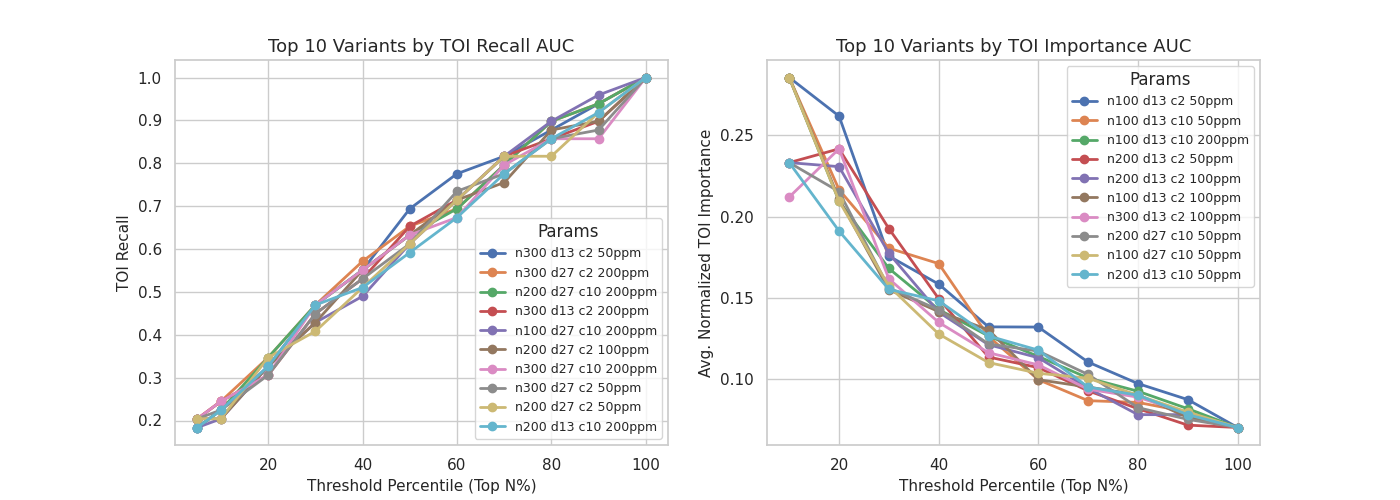}
    \caption{Performance of top 10 URF-4 variants across anomaly threshold percentiles. Left panel shows TOI Recall AUC curves for the highest-performing parameter configurations. Right panel presents TOI Importance AUC trajectories.}

    \label{fig:top10_variants_performance}
\end{figure*}
Figure~\ref{fig:top10_variants_performance} presents the performance trajectories of the top 10 URF-4 variants ranked by TOI Recall AUC across varying anomaly threshold percentiles. It shows the top URF-4 optimal models achieving ~40\% recall in the top 25\% of anomaly scores. However, while \textbf{individual feature importance rankings guide parameter prioritization for focused optimization} (\ref{sec:astrophysicalrelevanceandmorphologicalclusteringofurfanomaliesfortransitdetection}), the combined metric analysis \ref{sec:combinedmetricanalysis1} becomes essential for interpreting current results since different parameters optimize different performance aspects. The \textbf{divergent trajectories between TOI Recall AUC and TOI Importance AUC} in Figure \ref{fig:top10_variants_performance} highlight the need for weighted combination scoring to balance detection completeness against astrophysical relevance in URF-4 variant selection. 

\subsection{Results: Generalization Evaluation of Balanced URF Variant ($\alpha$=0.5)}
\label{sec:res_generalization_alpha_p5}

Evaluation across 10 random subsets ($\approx$40,000 total light curves) yields Table \ref{tab:generalization_results}. TOI Recall AUC deviates only 0.014 below target, indicating consistent TOI prioritization across the anomaly score distribution. TOI Importance AUC exceeds the target by 0.023, suggesting the model performs even better than expected at detecting astrophysically significant targets. The binary recall of 32.1\% and anomaly rate of 27.9\% indicate closely related TOI Recall and Anomaly Rate as predicted by Figure \ref{fig:synthparamsvsrecallanomalyrate1}. The value of these metrics will be improved by predicting over a finer mesh of synthetic design features and using a more comprehensive model scoring metric as presented in Section \ref{sec:finermappingsec2a}. \newline

\begin{table}
\centering
\begin{tabular}{lccc}
\hline
Metric & Value & Target & Deviation \\
\hline
Mean TOI Recall AUC & 0.547 & 0.561 & -0.014 \\
Mean TOI Importance AUC & 0.143 & 0.120 & +0.023 \\
Mean Binary TOI Recall & 0.321 &  -  &  -  \\
Mean Anomaly Rate & 0.279 &  -  &  -  \\
\hline
\end{tabular}
\caption{Generalization performance of balanced URF-4 variant ($\alpha$ = 0.5)}
\label{tab:generalization_results}
\end{table}

\textbf{Verdict}: This model satisfies our generalization criteria (that it will perform consistently on similar data), with both metrics falling within the acceptable ±0.03-0.05 deviation range.

\begin{table*}
\centering
\begin{tabular}{lcccc}
\hline
\textbf{Metric} & \textbf{$\alpha$ = 0.3} & \textbf{$\alpha$ = 0.5} & \textbf{$\alpha$ = 0.9} & \textbf{Best Performer} \\
\hline
TOI Recall AUC & 0.521 & 0.547 & \textbf{0.557} & $\alpha$ = 0.9 \\
TOI Importance AUC & 0.141 & \textbf{0.143} & 0.139 & $\alpha$ = 0.5 \\
Mean TOI Importance (Top 20\%) & 0.393 & \textbf{0.422} & 0.359 & $\alpha$ = 0.5 \\
Binary TOI Recall & 0.493 & 0.321 & \textbf{0.682} & $\alpha$ = 0.9 \\
Anomaly Rate & 0.447 & \textbf{0.279} & 0.558 & $\alpha$ = 0.5 \\
Top 10\% Importance TOIs – Score Percentile Range & \textbf{1.6–9.7\%} & 4.9–10.6\% & 2.0–10.2\% & $\alpha$ = 0.3 \\
Top 20\% Importance TOIs – Score Percentile Range & \textbf{1.3–28.4\%} & 3.1–24.4\% & 1.4–23.8\% & $\alpha$ = 0.3 \\
\hline
\end{tabular}

\caption{Performance comparison of URF-4 $\alpha$ variants across six evaluation metrics}
\label{tab:alpha_comparison}
\end{table*}

\begin{table*}
\centering
\begin{tabular}{lcccccc}
\hline
Metric & S1 Best & S2 Best & S3 Best & S4 Best & S5 Best & Consistency \\
\hline
TOI Recall AUC & $\alpha$=0.5 & $\alpha$=0.9 & $\alpha$=0.9 & $\alpha$=0.9 & $\alpha$=0.5 & 60\% $\alpha$=0.9 \\
TOI Importance AUC & $\alpha$=0.9 & $\alpha$=0.5 & $\alpha$=0.5 & $\alpha$=0.3 & $\alpha$=0.3 & Mixed \\
TOI Importance Top 20\% & $\alpha$=0.5 & $\alpha$=0.5 & $\alpha$=0.5 & $\alpha$=0.5 & $\alpha$=0.5 & 100\% $\alpha$=0.5 \\
Binary TOI Recall & $\alpha$=0.9 & $\alpha$=0.9 & $\alpha$=0.9 & $\alpha$=0.9 & $\alpha$=0.9 & 100\% $\alpha$=0.9 \\
Anomaly Rate (lowest) & $\alpha$=0.5 & $\alpha$=0.5 & $\alpha$=0.5 & $\alpha$=0.5 & $\alpha$=0.5 & 100\% $\alpha$=0.5 \\
Top 10P Importance Range& $\alpha$=0.3 & $\alpha$=0.3 & $\alpha$=0.3 & $\alpha$=0.5 & $\alpha$=0.5 & 60\% $\alpha$=0.3 \\
Top 20P Importance Range& $\alpha$=0.3 & $\alpha$=0.9 & $\alpha$=0.5 & $\alpha$=0.5 & $\alpha$=0.9 & Mixed \\
\hline
\end{tabular}
\caption{Cross-sector performance of $\alpha$ variants across TESS Sectors 1-5}
\label{tab:crosssectoralphavariantperformance}

\end{table*}

\subsubsection{Results: Alpha Variant Performance Comparison}{} 
Evaluation across 10 consistent test subsets (40,000 total light curves per model) reveals clear and reproducible performance differences between $\alpha$ variants, as summarized in Table~\ref{tab:alpha_comparison}. \newline

\subsubsection{Intepretation of Results of Table \ref{tab:alpha_comparison}}
\label{sec:interpretationofalphacomparisonacrosssixmetricssamesector}

\textit{Behavioral Analysis: Systematic $\alpha$-Dependent Performance} - The three $\alpha$-variant models demonstrate \textbf{consistent and distinct behavioral patterns} that align with their design objectives: \newline

\textbf{$\alpha$ = 0.9 (High Recall Priority):} Emphasizes \textbf{detection completeness}, achieving the highest TOI Recall AUC (0.557) and Binary TOI Recall (68.2\%). However, this comes at the cost of an elevated anomaly rate (55.8\%), indicating increased false positive burden for follow-up observations. \newline

\textbf{$\alpha$ = 0.3 (High Importance Priority):} Optimizes \textbf{early ranking quality}, concentrating the most scientifically valuable TOIs into narrow score percentile ranges (top 10\% importance TOIs span only 1.6–9.7\% of score space). This model achieves effective importance-based prioritization while maintaining moderate recall performance. \newline

\textbf{$\alpha$ = 0.5 (Balanced Priority):} Provides an \textbf{optimal trade-off}, achieving the lowest anomaly rate (27.9\%) while maintaining strong performance across importance metrics. The Mean TOI Importance in the top 20\% reaches 0.422, indicating effective early detection of high-value targets with minimal false positive contamination. \newline

These results \textbf{validate the central hypothesis}: synthetic set design successfully steers model behavior toward different scientific objectives. Each $\alpha$ variant exhibits predictable performance characteristics that remain stable across randomized subsets, demonstrating that:

\begin{enumerate}
    \item \textbf{$\alpha$-parameter control works as designed}
    \item \textbf{Behavioral differences are robust}
    \item \textbf{Trade-offs are quantifiable} \end{enumerate}

\subsubsection{Results: Cross-Sector Generalization of Alpha Variants}
\label{sec:res_cross_sector_gen_alpha_var}

Cross-sector evaluation across TESS Sectors 1-5 demonstrates \textbf{robust generalization} of $\alpha$-dependent behavioral patterns, validating the transferability of synthetic training across diverse stellar populations.

\textbf{Performance Consistency:} Each $\alpha$ variant maintains its designed priorities across all sectors:
\begin{itemize}
    \item \textbf{$\alpha$ = 0.9}: Achieves best Binary TOI Recall in \textbf{all 5 sectors} (100\% consistency) with \textbf{Mean Recall} ranging from \textbf{0.509} to \textbf{0.762} and best TOI Recall AUC in 3/5 sectors (60\% consistency).
    \item \textbf{$\alpha$ = 0.3}: Wins Top 10\% Importance Range in 3/5 sectors (60\% consistency), consistently concentrating high-value targets in narrow score ranges
    \item \textbf{$\alpha$ = 0.5}: Achieves best TOI Importance Top 20\% and lowest Anomaly Rate in \textbf{all 5 sectors} (100\% consistency each) with \textbf{TOI Importance Top 20P} ranging from \textbf{0.299} to \textbf{0.486} and Anomaly Rate ranging from \textbf{0.269} to \textbf{0.311} across the five sectors.
\end{itemize}

\textbf{Operational Implications:} The predictable cross-sector behavior enables confident deployment of $\alpha$ variants to new TESS sectors without retraining.

\subsection{Results: t-SNE Clustering on Stellar Parameters with N-WRSS overlay}

\label{sec:stellar_tsne_results}

\begin{figure}
    \centering
    \includegraphics[width=1\linewidth]{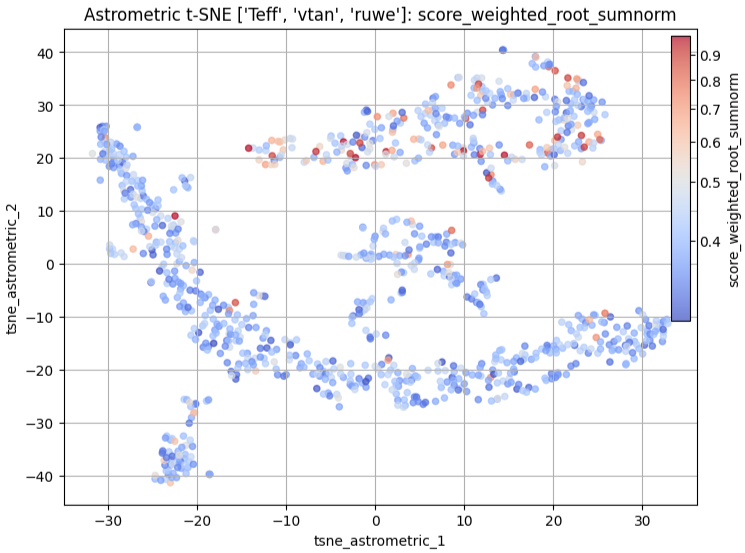}
    \caption{t-SNE clustering of Sector 1 N-WRSS $\ge$ 0.6 curves based on [Teff, vtan, ruwe]. N-WRSS goes from blue to red increasing from 0.0 to 0.9 (non-linearly to mark N-WRSS $\ge$ 0.6 better visually)}
    \label{fig:teffvtanruwesec1highnwrss}
\end{figure}

The most visually separable and physically interpretable feature set was: \textbf{['Teff', 'vtan', 'ruwe']}, showing excellent concentrated red region (high N-WRSS) clusters. These three features showed up consistently across higher silhouette rankings of feature combinations. Testing it on sector 1 data showed the t-SNE clustering as visualized in Figure \ref{fig:teffvtanruwesec1highnwrss}

\begin{figure}
    \centering
    \includegraphics[width=1\linewidth]{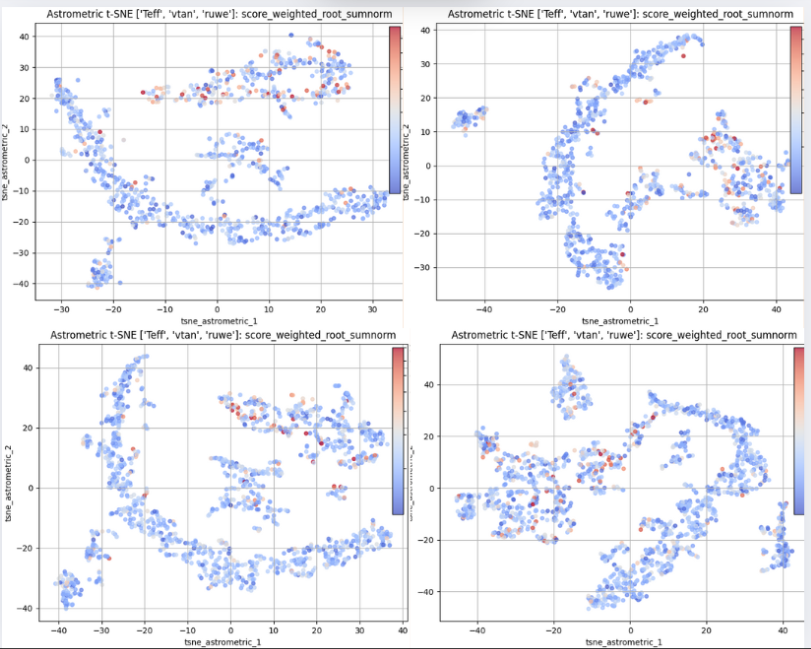}
    \caption{SECTOR 1 - 4 (CLOCKWISE FROM TOP LEFT) on feature set TEFF VTAN RUWE with N-WRSS on same color scale as Figure \ref{fig:teffvtanruwesec1highnwrss}}
    \label{fig:teffvtanruwecrosssectortsne}
\end{figure}

We performed cross sector t-SNE tests across TESS SPOC Sectors 1-4 based on the feature set [$T_{eff}$, $v_{tan}$, RUWE]. The t-SNE embeddings in Figure~\ref{fig:teffvtanruwecrosssectortsne} reveal consistent structural patterns across TESS Sectors 1–4.

The projection shows a recurring concentration of high N-WRSS scores (red regions) along narrow manifolds in the reduced space, suggesting the existence of a physically distinct anomaly subpopulation. This reinforces the hypothesis that high-scoring anomalies are not randomly distributed but instead correlate with specific stellar or kinematic regimes, validating the physical relevance of the URF-derived anomaly scores across sectors.

\begin{figure*}
    \centering
    \includegraphics[width=1\linewidth]{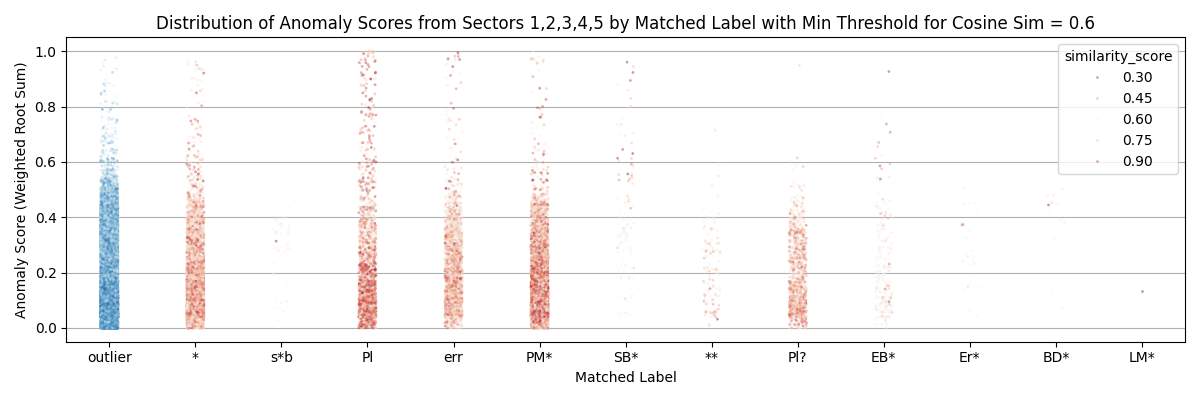}
    \caption{Distribution of N-WRSS from Sectors 1 - 5 by Matched Label from Cosine Similarity with a threshold of 0.6 for similarity score}
    \label{fig:cosine_all}
\end{figure*}

\begin{figure}
    \centering
    \includegraphics[width=1\linewidth]{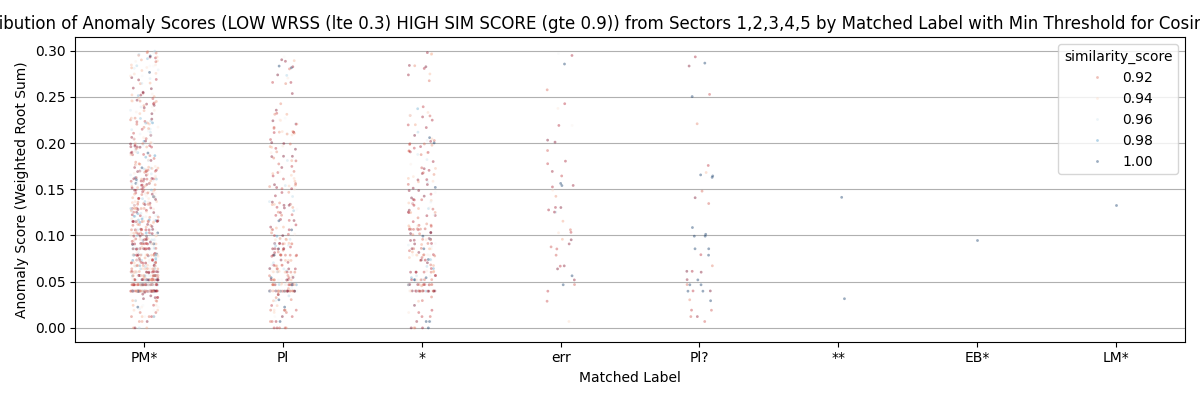}
    \caption{Distribution of Low N-WRSS and High Similarity Score from Sectors 1 - 5 by Matched Label from Cosine Similarity with a threshold of 0.6 for similarity score.}
    \label{fig:cosine_lowwrss_highsim}
\end{figure}
\begin{figure}
    \centering
    \includegraphics[width=1\linewidth]{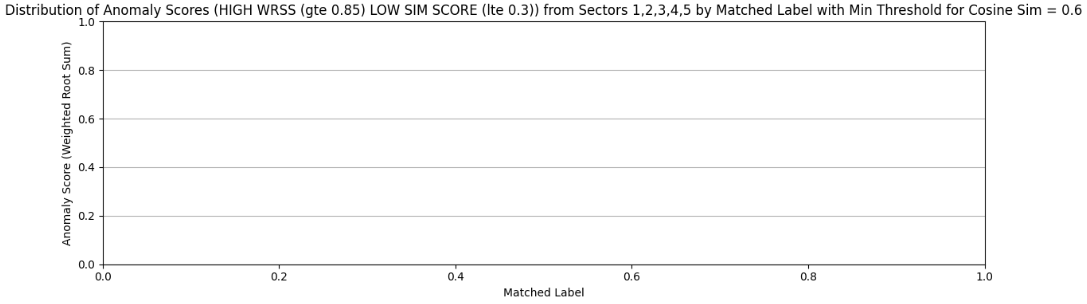}
    \caption{Distribution of High N-WRSS and Low Similarity Score from Sectors 1 - 5 by Matched Label from Cosine Similarity with a threshold of 0.6 for similarity score (\textbf{shows no curves with high anomaly score that could not be successfully classified into one of the SIMBAD Object type based morphology labels for the light curves})}
    \label{fig:cosine_highwrss_lowsim}
\end{figure}
\subsection{Results of Morphological Classification via Cosine Similarity}
\label{sec:cosinesimilaritymatchingresults}

To assess the astrophysical plausibility of morphologically matched labels, we analyze the distribution of URF anomaly scores across the assigned SIMBAD (\citet{SIMBAD}) label categories.

We visualize the results via strip plots of the N-WRSS against matched SIMBAD labels. To interpret these distributions, we define three separate diagnostic views:

\begin{enumerate}
    \item \textbf{Full distribution (Figure~\ref{fig:cosine_all})}: All matched candidates, regardless of anomaly score, are shown.

    \item \textbf{High similarity \& low anomaly score (Figure~\ref{fig:cosine_lowwrss_highsim})}: We restrict to candidates with N-WRSS $\leq 0.3$ and cosine similarity $\geq 0.9$. 

    \item \textbf{High anomaly score \& low similarity (Figure~\ref{fig:cosine_highwrss_lowsim})}: This diagnostic isolates candidates with anomaly scores $\geq 0.85$ and cosine similarity $\leq 0.3$.
\end{enumerate}

\subsubsection{Interpretation of Cosine Similarity Results}
\label{sec:cosine_similarity_interpretation}

The three diagnostic views (figures \ref{fig:cosine_all}, \ref{fig:cosine_lowwrss_highsim} and \ref{fig:cosine_highwrss_lowsim}) enable a structured assessment of how anomaly scores and morphological similarity align with known astrophysical classes. \newline

First, the full distribution (Figure~\ref{fig:cosine_all}) confirms that the majority of anomalies are morphologically consistent with stellar variability or planetary candidates  -  with concentrated clusters in \texttt{PM*}, \texttt{Pl}, and \texttt{*} label groups and \textbf{a significant population}, with almost constant density across N-WRSS scores, \textbf{of Planet-like label group above a threshold of N-WRSS=0.6 where other label groups have a much lesser and more scarce population}. \newline

Second, the high-similarity/low-anomaly set (Figure~\ref{fig:cosine_lowwrss_highsim}) demonstrates that \textbf{URF correctly assigns low anomaly scores to morphologically standard light curves which can be classified into a labeled category}. This supports the argument that low URF scores represent “normal” behavior in the training distribution. \newline

Finally, and most critically, the high-anomaly/low-similarity view (Figure~\ref{fig:cosine_highwrss_lowsim}) yields zero candidates. This strongly indicates that \textbf{nearly all high-score candidates still retain morphological similarity to known astrophysical classes}  -  and that the pipeline’s anomaly scores are physically meaningful rather than driven by noise or artifacts.

Taken together, these plots show that cosine similarity provides astrophysically grounded, interpretable labels to anomaly candidates  -  and that URF scores correlate well with label group structure in a physically plausible way.  \newline

In total, we processed \textbf{87,858} light curves spanning TESS SPOC Sectors 1 through 5. This forms the complete set of candidates from which anomaly scores were computed, morphological features extracted, and cosine similarity-based label assignments performed.

\begin{table}
\centering
\caption{Comparison of clustering methods applied to the 10-dimensional feature set extracted from URF-4 planet-like candidates. Each entry lists the number of TOIs, total curves, and the percentage of TOIs in that cluster.}
\label{tab:clustering_summary}
\begin{tabular}{lllll}
\hline
\textbf{Clustering} & \textbf{Label} & \textbf{TOIs} & \textbf{Curves} & \textbf{(\%)}\\
\hline
\multirow{3}{*}{HDBSCAN} 
    & Cluster -1 & 3 & 155 & 1.9 \\
    & Cluster 0  & 16 & 263 & 6.1 \\
    & Cluster 1  & 0 & 35  & 0.0 \\
\hline
\multirow{3}{*}{KMeans (k=3)} 
    & Cluster 0 & 11 & 228 & 4.8 \\
    & Cluster 1 & 1 & 144 & 0.7 \\
    & Cluster 2 & 7 & 81  & 8.6 \\
\hline
\multirow{7}{*}{DBSCAN} 
    & Cluster -1 & 2 & 68  & 2.9 \\
    & Cluster 0  & 16 & 304 & 5.3 \\
    & Cluster 1  & 1 & 57  & 1.8 \\
    & Cluster 2  & 0 & 5   & 0.0 \\
    & Cluster 3  & 0 & 7   & 0.0 \\
    & Cluster 4  & 0 & 5   & 0.0 \\
    & Cluster 5  & 0 & 4   & 0.0 \\
    & Cluster 6  & 0 & 3   & 0.0 \\
\hline
\multirow{2}{*}{\textbf{GMM (n=2)}} 
    & \textbf{Cluster 0} & \textbf{17} & \textbf{155} & \textbf{11.0} \\
    & Cluster 1 & 2 & 298 & 0.7 \\
\hline
\multirow{3}{*}{\textbf{DPMM (n=3)}} 
    & Cluster 0 & 2  & 175 & 1.1 \\
    & Cluster 1 & 1  & 164 & 0.6 \\
    & \textbf{Cluster 2} & \textbf{16} & \textbf{114} & \textbf{14.0} \\
\hline
\end{tabular}
\end{table}

\subsubsection{Results and Interpretation of Morphological Clustering Outcomes}
\label{sec:clustering_results_interp}

To explore the morphological diversity among the 620 planet-like candidates (Section \ref{sec:clusteringandmorphologicalfilteringmethods}), we applied a range of clustering algorithms to a 10-dimensional feature set (Section \ref{sec:curve_feature_2}) engineered from phase-folded light curves. This feature set is as described in section \ref{sec:curve_feature_2}. Our objective was to determine whether these features could reveal structural subgroups and isolate known TOIs within well-defined clusters. Results of clustering methods are shown in Table \ref{tab:clustering_summary}

Among all tested methods, \textbf{Gaussian Mixture Models (GMM)} \citet{Dempster1977} with two components produced the clearest separation: one cluster contained \textbf{17 TOIs among 155 curves}, achieving the highest TOI density. \textbf{Dirichlet Process GMM (DPMM)} \citet{Gorur2010} performed comparably well, grouping \textbf{16 TOIs within a smaller cluster of 114 curves}, indicating a stronger enrichment but with more fragmentation overall.

\begin{figure}
    \centering
    \includegraphics[width=1\linewidth]{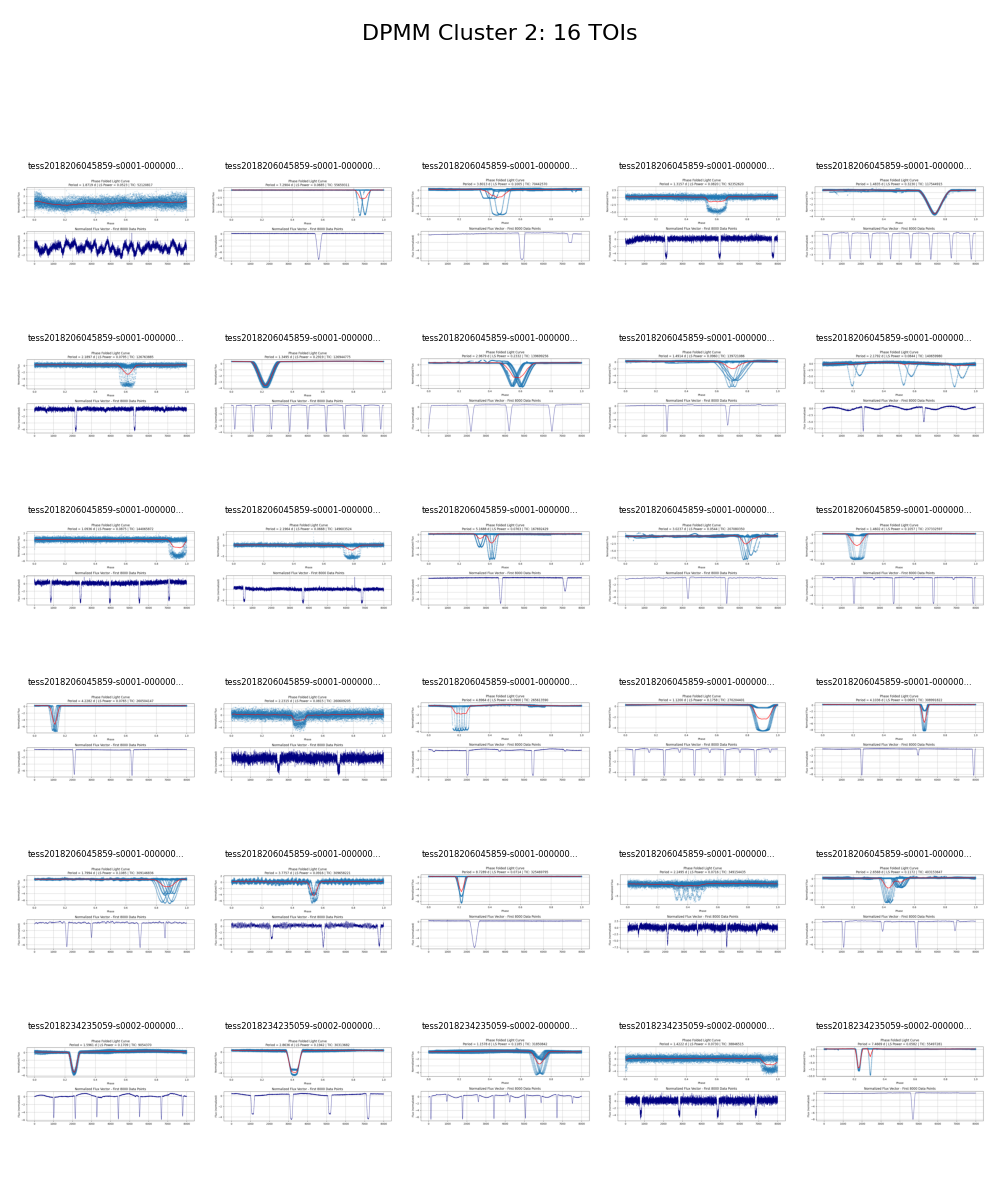}
    \caption{Random sample of curves from dpmm cluster 2}
    \label{fig:dpmmc2sample}
\end{figure}

\begin{figure}
    \centering
    \includegraphics[width=1\linewidth]{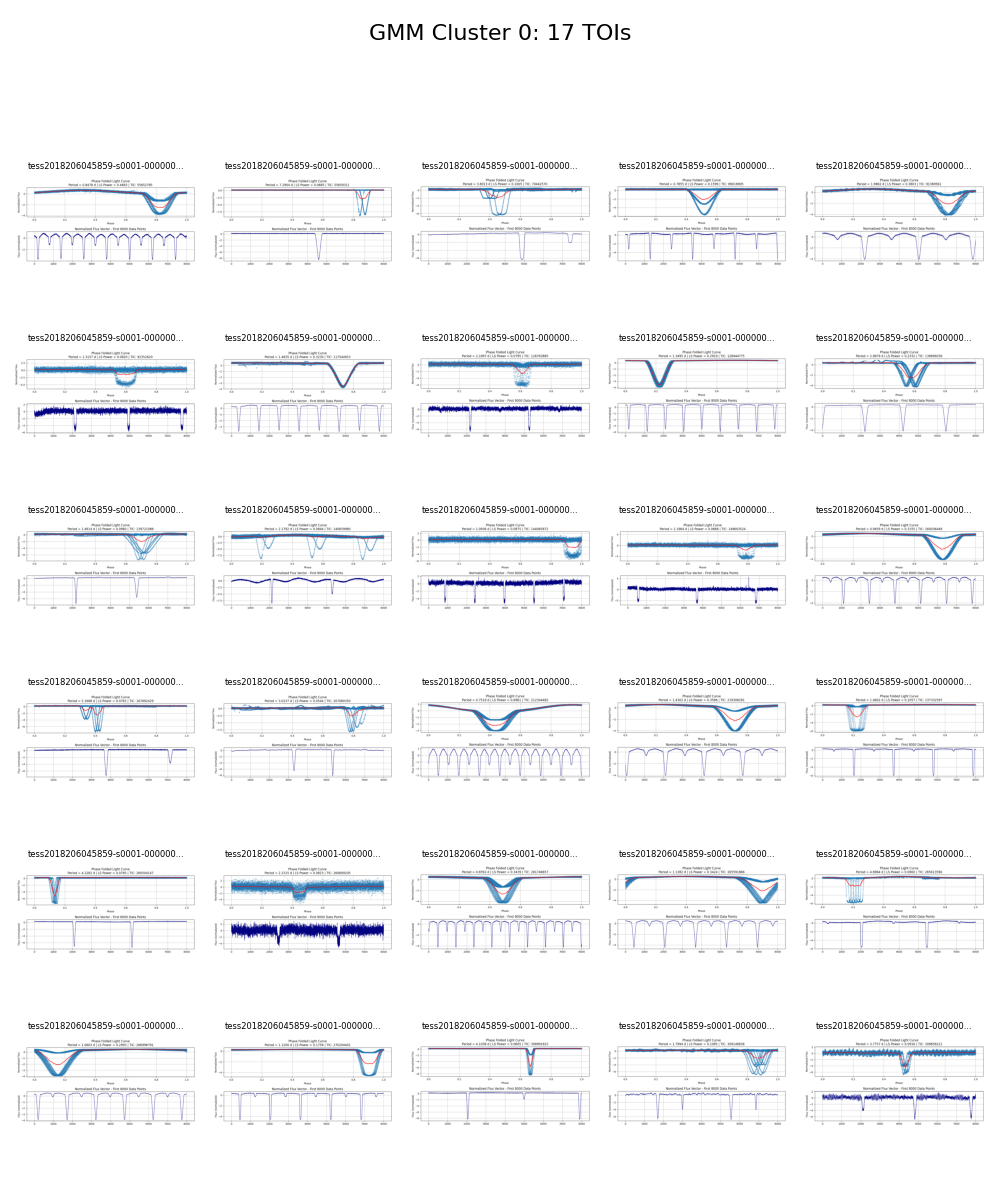}
    \caption{Random sample of curves from gmm cluster 0}
    \label{fig:gmmc0sample}
\end{figure}

While other methods such as DBSCAN, HDBSCAN, and KMeans \citet{Fotopoulou_2024} provided partial clustering success, they either included fewer TOIs or distributed them across broader clusters. The probabilistic nature of GMM and DPMM allowed them to better model the subtle variations in transit morphology, making them more suitable for this task. As evident from Figures \ref{fig:dpmmc2sample} and \ref{fig:gmmc0sample}, the planet like candidate list of 620 curves with $N-WRSS \geq 0.35$ (arbitrary threshold) filtered down to a maximum of 155 curves with morphology matching the curves of the synthetic training set used for fine tuning hyper-parameters, 17 of which are TESS Objects of Interest as listed in Table \ref{tab:dpmmc2tois} and Table \ref{tab:gmmc0tois}.

\textbf{In conclusion}, the combination of interpretable features and flexible model-based clustering yielded the best separation of high-quality transit-like signals. The GMM and DPMM results (sampled in figures \ref{fig:dpmmc2sample} and \ref{fig:gmmc0sample} to visualize morphological similarity) represent the most promising configurations to guide further candidate filtering, validation, and prioritization.

\subsection{Initial Injection-Recovery Test Results}
\label{sec:init_inj_rec_results}

Before performing the Monte Carlo simulations of the injection-recovery test, we first injection-recovery tested each of the 10 configurations with 1980 randomly sampled real TESS curves and the 20 curves of that particular configuration for each configuration test set. The clear tradeoff trends are visible across N-WRSS thresholds in the results of this test as presented in Table \ref{tab:init_inj_rec_summary}. The $N-WRSS \ge 0.35$ achieves near-perfect recall but at the expense of $\approx 126$ candidates to vet per 20 injections, serving as an intentionally permissive baseline. Conversely, $N-WRSS \ge 0.6$ achieves an average of 84\% recall with $\approx 24$ curves to vet, representing a more practical operating point. At $N-WRSS \ge 0.9$ precision is high but recall falls sharply, illustrating the cost of overly strict thresholds.

\begin{table}
\centering
\caption{Average recovery performance in the initial injection–recovery test. Detailed summary are present in Tables \ref{tab:injrec_thr035}, \ref{tab:injrec_thr060} and \ref{tab:injrec_thr090}.}
\label{tab:init_inj_rec_summary}
\begin{tabular}{lcc}
\hline
N-WRSS Threshold& Avg. Retrieved & Avg. Total \\
\hline
0.35 & 19.6 / 20 & 126 \\
0.60 & 16.7 / 20 & 24 \\
0.90 & 5.3 / 20  & 5 \\
\hline
\end{tabular}
\end{table}

\subsection{Evaluation of Injection-Recovery Performance}
\label{sec:mc_eval_metrics}

\begin{table}
\centering
\caption{Monte Carlo injection–recovery performance over 80 runs at $N-WRSS\geq 0.6$. Reported values are mean $\pm$ standard deviation across runs, with 95\% confidence intervals on the mean in parentheses.}
\label{tab:mc_injrec_summary}
\begin{tabular}{lcc}
\hline
Metric & Mean $\pm$ SD & 95\% CI on Mean \\
\hline
Precision & $0.53 \pm 0.18$ & [0.49, 0.57] \\
Recall & $0.74 \pm 0.24$ & [0.69, 0.79] \\
Retrieved Injected & $14.8 \pm 4.8$ & [13.8, 15.9] \\
Retrieved Total & $26.1 \pm 8.0$ & [24.4, 27.9] \\
\hline
\end{tabular}
\end{table}

\begin{figure*}
    \centering
    \includegraphics[width=\textwidth]{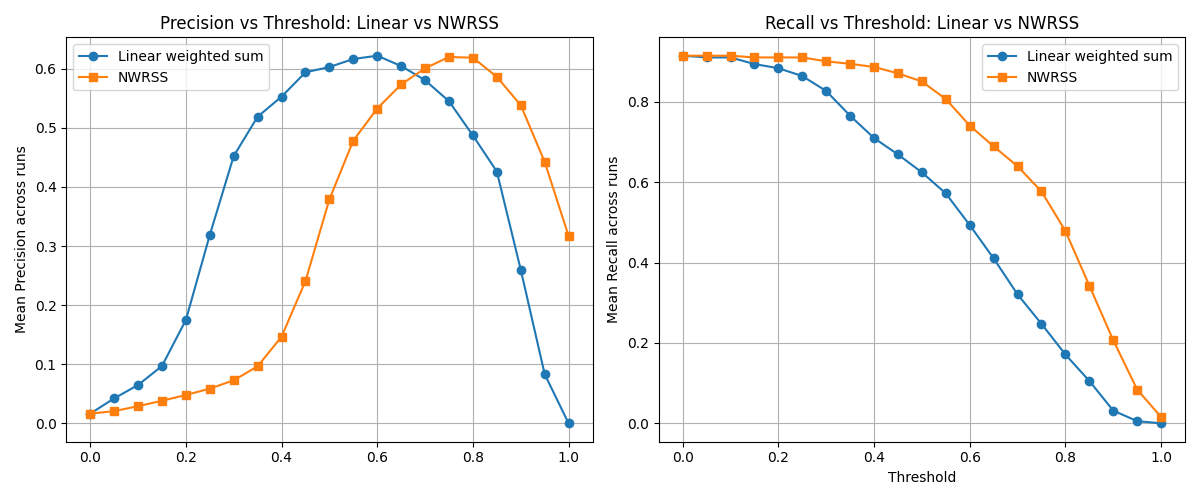}
    \caption{Comparison of precision (left) and recall (right) as functions of the detection threshold for the linear weighted sum and the normalized weighted root sum score (N-WRSS) in terms of measuring model consesus towards anomalousness. N-WRSS maintains higher recall at stricter thresholds while achieving improved precision in the high-threshold regime, indicating greater robustness against false positives.}
    \label{fig:linear_vs_nwrss}
\end{figure*}

To more rigorously quantify performance and mitigate dependence on any single injection set, we conducted over 80 Monte Carlo runs (Section \ref{sec:mc_setup}) of the injection-recovery test using randomly sampled TESS light curves and synthetic transits. At a practical operating point of $N-WRSS \ge 0.6$, the ensemble statistics across runs (Table \ref{tab:mc_injrec_summary}) yield a mean precision of $0.53 \pm 0.18$ and mean recall of $0.74 \pm 0.24$. 95\% confidence intervals on these metrics were $0.49-0.57$ and $0.68-0.79$ respectively. On average, 14.8 $\pm$ 4.8 injected curves were successfully recovered per run, with $26.1 \pm 8.0$ total candidates requiring vetting (95\% confidence intervals: $13.8-15.9$ recovered and $24.4-47.9$ to vet). These results reinforce the earlier tradeoff trends, demonstrating that the chosen threshold balances high recall with a manageable number of candidates, while also confirming robustness across repeated randomized trials. A full visualization of performance across all thresholds is provided in Appendix~\ref{fig:mc_injrec_thresholdsweep}. \newline

To further validate the choice of N-WRSS over a simpler linear weighted sum for ensemble consensus scoring (see Section \ref{sec:wrss}), we compared both strategies across 80+ Monte Carlo runs. Figure \ref{fig:linear_vs_nwrss}
 shows mean precision and recall as functions of detection threshold. The linear weighted sum yields a sharp precision peak near 0.6,  but at the cost of collapsing recall, underscoring its sensitivity to threshold choice. In contrast, N-WRSS provides  smoother, more stable tradeoffs, maintaining a competitive precision while sustaining higher recall across a wider threshold range. This robustness complements its mathematical basis as a monotone concave transformation that suppresses single-model dominance in ensemble scoring.

\begin{figure*}
    \centering
    \includegraphics[width=1\linewidth]{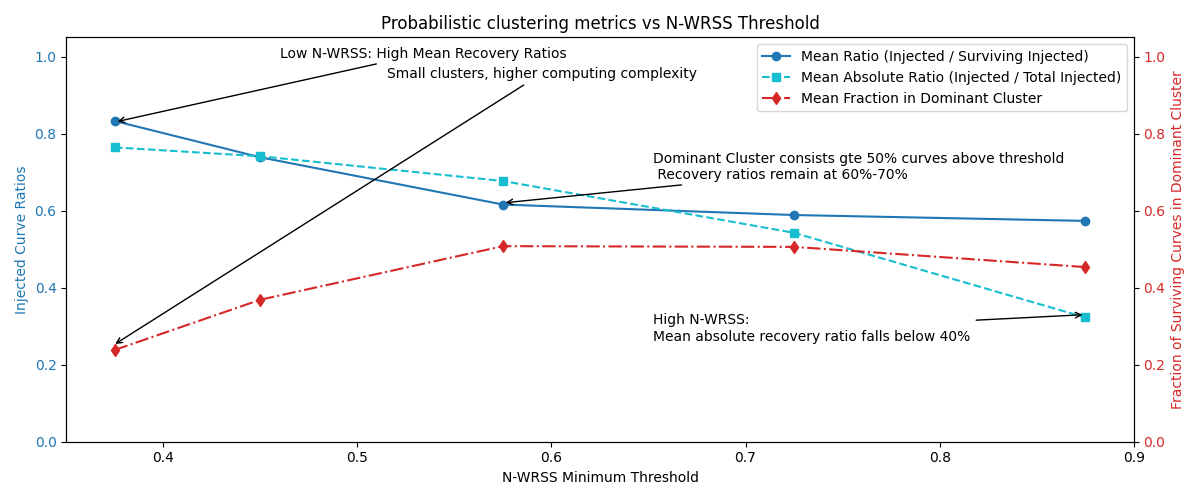}
    \caption{Quantification of URF steering across $N$-WRSS thresholds. Shown are mean recovery ratios, dominant cluster fractions, and absolute recovery ratios from Monte Carlo injection–recovery experiments. At intermediate thresholds ($0.5 \lesssim N$-WRSS $\lesssim 0.7$), injected transits are preferentially steered into a single dominant cluster while maintaining $\sim$60--70\% recovery, whereas stricter thresholds suppress both false positives and true signals. Numerical values are provided in Table~\ref{tab:urf_steering}.}
    \label{fig:urf_steering}
\end{figure*}

The injection-recovery tests allow us to quantify the clustering behavior of an unsupervised probabilistic clustering method like the Gaussian Mixture Model within a Dirichlet process framework. Figure \ref{fig:urf_steering}
 and Table \ref{tab:urf_steering} summarize three key statistics:

 \begin{enumerate}
     \item Mean ratio of injected curves in the dominant cluster relative to all injected curves surviving the threshold.
     \item The mean absolute recovery ratio. This is the number of injected curves in the dominant cluster relative to the total number of injected curves (20) regardless of N-WRSS threshold.
     \item The mean fraction of surviving curves assigned to the dominant cluster. This is the number of curves in the dominant cluster relative to the total number of curves that survive an N-WRSS threshold.

 \end{enumerate}
 At permissive thresholds ($N-WRSS$ threshold $\leq 0.35$), recovery ratios approach unity, forming small clusters (20\% of curves that survive the threshold per cluster), with high recovery ratios (75-80\% injected curves in this single cluster), resulting in high computational complexity. As thresholds increase ($0.5 \lesssim $ N-WRSS $\lesssim 0.7$), injected curves are steered into a single dominant cluster comprising $\gtrsim 50\%$ of all surviving curves, with mean recovery ratios sustained at 60-70\%. At overly strict thresholds ($N-WRSS \gtrsim 0.8$), the mean absolute recovery ratio falls below 40\%, reflecting the significant loss of true signals alongside the suppression of spurious clusters. This analysis demonstrates that the URF does not simply filter anomalies but actively steers recovered injections toward coherent morphological groupings. This provides an unsupervised basis for distinguishing transit-like morphological signals from noise and empirically justifies the use of probabilistic clustering on a morphological feature set. 

 The false positives, as presented in Table \ref{tab:simbad_fp}, are predominantly explained by well-known astrophysical contaminants. Out of 187 non-injected curves that surpass the N-WRSS threshold and are in the dominant cluster, the largest contribution arises from binaries (100 objects), consistent with eclipsing binary light curves mimicking transit-like morphology. Generic stars (50) and variables of rotational, pulsational, or eruptive type (20) account for most of the remaining population. A small but non-negligible set of objects classified as planets or planet-like (8) demonstrates that non-injected but genuine planet-hosting curves do surpass the threshold. Finally, unresolved multiples (6) and unclassified sources (3) constitute a minor fraction. These results indicate that at the practical operating point of $N-WRSS \geq 0.6$, CLARA's false positives are astrophysically interpretable and dominated by binaries and stellar variability.

\begin{table*}
\centering
\caption{Quantification of probabilistic clustering on morphological feature set across $N$-WRSS threshold ranges. Reported values are mean statistics across Monte Carlo runs. The Mean Ratio refers to injected curves surviving in the dominant cluster relative to all the injected curves that survive an N-WRSS threshold; the Mean Absolute Ratio is relative to the total number of injected curves (20); the Mean Fraction in Dominant Cluster is the fraction of all surviving curves assigned to the dominant cluster. The dominant cluster is the one containing the most injected curves above an N-WRSS threshold}
\label{tab:urf_steering}
\begin{tabular}{lccccc}
\hline
$N$-WRSS Range & Mean CI Lower & Mean CI Upper & Mean Ratio & Mean Fraction in Dominant Cluster & Mean Absolute Ratio \\
\hline
0.35--0.40 & 0.51 & 1.00 & 0.83 & 0.24 & 0.76 \\
0.40--0.50 & 0.42 & 0.98 & 0.74 & 0.37 & 0.74 \\
0.50--0.65 & 0.37 & 0.86 & 0.62 & 0.51 & 0.68 \\
0.65--0.80 & 0.37 & 0.97 & 0.59 & 0.51 & 0.54 \\
0.80--0.95 & 0.30 & 0.97 & 0.57 & 0.45 & 0.32 \\
\hline
\end{tabular}
\end{table*}

 \begin{table}
\centering
\caption{SIMBAD classifications of false positives (non-injected curves in dominant cluster with N-WRSS $\geq 0.6$)  from Monte Carlo injection-recovery runs. Objects are grouped into broad astrophysical categories.}
\label{tab:simbad_fp}
\begin{tabular}{lc}
\hline
Category & Count \\
\hline
Binaries & 100 \\
Variables (rot./puls./eruptive) & 20 \\
Generic Stars & 50 \\
Planets or Planet-like & 8 \\
Multiple/Unresolved systems & 6 \\
Other/Unclassified & 3 \\
\hline
Total & 187 \\
\hline
\end{tabular}
\end{table}

\section{Discussion}

\subsection{Synthesis of Key Findings}

CLARA demonstrates that synthetic set design enables interpretable, tunable, and efficient URF-based anomaly detection in large astronomical surveys, linking model behavior directly to training parameters. Parameter sweeps over 36 URF-4 variants reveal transit duration as the dominant driver of TOI Recall and Anomaly Rate, and noise level as the key factor in importance-based prioritization, enabling predictive tuning within 50–300\,ppm noise and 13–27\,day duration ranges. The $\alpha$-parameterized scoring framework robustly controls completeness–significance trade-offs across TESS sectors, with $\alpha=0.9$ maximizing TOI recall and $\alpha=0.3$ concentrating high-value targets; combined metrics like N-WRSS unify their strengths.

\subsection{Astrophysical Validation and Physical Interpretability}

Anomaly–stellar parameter correlations and t-SNE embeddings show high-scoring curves form coherent astrophysical manifolds. SIMBAD cross-matching confirms that N-WRSS $\geq 0.85$ candidates always exhibit high morphological similarity to known classes, supporting physical authenticity. N-WRSS favors planet-like candidates at high thresholds, indicating appropriate sensitivity to transit morphologies.

\subsection{Limitations and Scope of Applicability}

Current results apply to 2-minute SPOC targets, with FFI data requiring recalibrated thresholds and parameter–performance mappings. Dependence on TOIs for validation risks bias, partly mitigated by SIMBAD-based checks and injection-recovery tests. Simplified synthetic models and truncated light curves may omit complex or long-duration anomalies, motivating more realistic generation and extended feature extraction.

\subsection{Future Directions and Extensions}
\label{sec:future_work}

CLARA enables seven future research avenues: \newline
(1) scale to a 500k light curve benchmark via \href{https://github.com/googleboy-byte/clara-cli}{clara-cli} to test full-pipeline scalability; \newline
(2) implement rigorous statistical validation with confidence intervals, cross-method comparisons, and deeper analysis of why Teff, RUWE, and vtan correlate with high N-WRSS; \newline
(3) expand morphological analysis with advanced feature combinations, deep-learned representations, and improved discovery metrics; \newline 
(4) adapt framework to other surveys (Kepler, LSST, K2) via modular synthetic generation; \newline 
(5) formalize N-WRSS theory, including square-root weighting justification, alternative fusion strategies, and a principled $\alpha$-controlled performance metric with Pareto optimization; \newline
(6) develop automated deep learning meta-optimization for synthetic parameter tuning; \newline
(7) integrate GAN-generated light curves for supervised post-clustering classification to filter false positives without TOI labels.

\section{Conclusions}

We present CLARA, a modular URF-based framework for unsupervised transit detection whose behavior is steered by synthetic training set design. Part I introduced URF-4, applied $\alpha$-dependent scoring to tune completeness ($\alpha=0.9$), significance ($\alpha=0.3$), or balance ($\alpha=0.5$), and validated consistent performance across 384k light curves from TESS Sectors 1–5. Part II used Gaia DR3 parameters, t-SNE embeddings, and SIMBAD morphological matching to confirm that N-WRSS–ranked anomalies occupy coherent astrophysical subspaces and align with known exoplanetary and stellar variability classes, with no high-score, low-similarity false positives. CLARA thus delivers a physically grounded, interpretable, and generalizable anomaly detection pipeline, extendable to other surveys and RV cross-matching.

\section{Data Availability}

\label{url:gitrepo}The full CLARA pipeline, including Jupyter notebooks, source code, pre-trained URF models, and helper scripts for light curve preprocessing, anomaly scoring, and similarity matching, and the CLARA cli tool to execute the pipeline with the computational resources and efficiency as stated in section \ref{sec:computeresourcesandruntime}, is publicly available at: \url{https://github.com/googleboy-byte/CLARA} and \url{https://github.com/googleboy-byte/clara-cli}. \newline

Intermediate results, including anomaly scores, N-WRSS scores, cosine similarity matches, and t-SNE cluster outputs, are provided as supplementary CSVs and figure files in the repository’s \texttt{data/} and \texttt{results/} directories. \newline

All TESS light curve data used in this study were obtained from the Mikulski Archive for Space Telescopes (MAST) at \url{https://archive.stsci.edu/tess/}. Specific sectors analyzed include SPOC 2-minute cadence data from TESS Sectors 1–5. \newline

TOI (TESS Object of Interest) catalogs were retrieved from the Exoplanet Follow-up Observing Program for TESS (ExoFOP-TESS): \url{https://exofop.ipac.caltech.edu/tess/download_toi.php?sort=toi&output=csv}

Stellar and astrometric parameters were retrieved from the TESS Input Catalog (TIC v8.2) and Gaia Data Release 3 (Gaia DR3), both accessible via the MAST and ESA Gaia Archive respectively. \newline

Labeled reference sets for cosine similarity matching were constructed from TOIs cross-matched with SIMBAD object classifications. SIMBAD data were obtained from the Centre de Données astronomiques de Strasbourg (CDS): \url{https://simbad.u-strasbg.fr/simbad/}. \newline

Synthetic light curves were generated using the \texttt{batman} transit model package (\url{https://github.com/lkreidberg/batman}). All synthetic parameter configurations, injection scripts, and random seeds used in this work are included in the GitHub repository to support full reproducibility. \newline

\begin{acknowledgments}
This research was conducted entirely independently and without institutional support or funding. All computations were performed on a personal desktop machine (Intel i3 processor, 32 GB RAM), which, despite its modest capacity, was sufficient for this work.

The author thanks the developers and maintainers of the open-source software and data archives acknowledged in the Data Availability section and Computational Resources (\ref{sec:computeresourcesandruntime})section, which made this work possible.
\end{acknowledgments}

\begin{contribution}

Mainak Dasgupta conceived the research problem, designed and implemented the CLARA methodology, performed the simulations and analysis, and wrote the manuscript. The codebase and documentation were also developed and maintained by the author.

\end{contribution}

%


\software{astropy, lightkurve, batman, scikit-learn}

\bibliography{main}{}
\bibliographystyle{aasjournalv7}


\appendix

\section{Computational Resources and Runtime}
\label{sec:computeresourcesandruntime}

All computations were performed on a personal desktop running Ubuntu 24.04 LTS (kernel 6.11), equipped with an Intel\textsuperscript{\textregistered} Core\texttrademark~i3-8100 CPU (4 physical cores, 3.60~GHz), 32~GB of DDR4 RAM, and no discrete GPU. The system used integrated Intel UHD Graphics 630 for all display tasks. CLARA was developed and executed entirely on this CPU-based machine without GPU acceleration. \newline

All code was written and executed in Jupyter notebooks launched using \texttt{python3 -m notebook} from within a Python virtual environment. The environment was configured with Python~3.12 and included standard scientific and astronomical computing libraries such as \texttt{numpy}, \texttt{pandas}, \texttt{astropy}, \texttt{scikit-learn}, \texttt{tqdm}, and \texttt{batman-package}. \newline

Parallel processing was enabled via Python’s \texttt{multiprocessing} module. All multi-threaded tasks - including light curve downloading, URF scoring, and cosine similarity classification - were executed using up to 4 parallel workers, matching the CPU's physical core count. \newline

Downloading $\sim$16{,}000 2-minute cadence SPOC light curves from a single TESS sector took approximately 18 minutes using 4 threads. Running the full anomaly scoring phase (URF-4) on the same sector took $\sim$1.5 hours per $\alpha$ model (e.g., $\alpha=0.3$ scored 15{,}878 light curves in 89 minutes; $\alpha=0.9$ in 86 minutes). Cosine similarity classification across $\sim$7{,}400 light curves took approximately 34 minutes using 4 workers. \newline

Additional preprocessing steps such as Lomb-Scargle computation, phase folding, and SIMBAD-based label similarity required $\sim$6 minutes for every 1{,}100 folded light curves. All tasks were executed on local storage spanning 3.7~TiB of disk capacity (including two external USB drives), with approximately 1.8~TiB utilized over the course of experimentation. \newline

\section{Notes}

\subsection{URF-3 Cross-Sector Validation}
\label{sec:noteonurf3crosssectorvalidation}

URF-3's performance exhibits significant sector-dependent variation due to its training methodology. Since URF-3 uses feature vectors extracted from confirmed TOIs in Sector 1 as its synthetic anomaly class, it demonstrates clear overfitting when evaluated on the same sector. URF-3's anomaly scores for TOIs in Sector 1 are systematically higher (peaking around 0.45-0.47 compared to a Non-TOI distribution peaking at 0.10-0.12) compared to Sector 2 (peaking around 0.10-0.12). The reduced score separation in Sector 2 reflects URF-3's diminished ability to generalize beyond the specific feature characteristics of its training TOIs. This sector dependency underscores the importance of independent validation sets and highlights a fundamental limitation of using confirmed target features directly as synthetic training data, reinforcing URF-4's advantage in using parametric synthetic generation for improved generalization.

\subsection{Hyper-parameter Search Space}
\label{sec:hyperparamsearchspaceappendix}

\verb|n_estimators|: 10 evenly spaced values between 50 and 200

\verb|max_features|: [sqrt, log2]

\verb|max_depth|: [100, 300, 500, 700, 900, 1000, None]

\verb|min_samples_split|: [2, 4, 7, 10]

\verb|min_samples_leaf|: [1, 2]

\verb|bootstrap|: [True, False]

\verb|warm_start|: [True, False]
\newline

\subsection{Note on Duration Units}
\label{sec:noteonduration}

During the course of this research, the duration of transit-like events is consistently measured in \textbf{days} across all computations, analyses, and visualizations. However, in a few instances, particularly in intermediate plots or variable names, the duration may be mistakenly labeled as \verb|duration_hr|. This is a \textbf{human oversight in labeling only}. All such instances refer to durations that are in fact expressed in \textbf{days}, not hours.

This clarification ensures consistency in interpretation and should be kept in mind when reviewing figures, tables, or variable definitions that include this mislabeled term.

\subsection{Note on TOI Importance AUC}
\label{sec:noteontoiimpauc}
While TOI Importance AUC varied enough in value to give us an opportunity to analyze trends of synthetic parameters v/s influence, more performance metrics, as discussed in table \ref{tab:alpha_metrics}, were developed to better represent quality and ranking of curves than TOI Importance AUC. To calculate AUC, we used the trapezoid function from the scipy python library.\newline 

\subsection{Note on Model Selection Strategy: Minimal Supervision Framework}
\label{sec:noteonmodelselectionstrategy}

The CLARA pipeline employs \textbf{unsupervised anomaly detection} using URFs trained with synthetic light curves as the outlier class. Crucially, URF models are \textbf{never exposed to labeled data with marked TOIs} during training or anomaly scoring.

However, to interpret and steer model behavior, we introduce a \textbf{light supervisory step} using a small, representative benchmark  -  a randomly sampled subset of ~4,000 real light curves from a sector (sector 2 in our case), enriched with known TOIs. This benchmark enables computation of an \textbf{$\alpha$-score} that summarizes each model's alignment with chosen scientific priorities.

Once performance trends are established across URF subvariants using this small labeled set, models selected via $\alpha$ demonstrate \textbf{robust and predictable behavior across larger datasets and different sectors}  -  despite never being trained on those datasets. This approach combines the scalability of unsupervised learning with the interpretability of minimal supervision, enabling researchers to build reliable, scientifically aligned anomaly detectors without requiring extensive labeling efforts.

\section{Astrometric and Stellar Features Description}
\label{sec:astrostellarfeaturedesc}

\begin{itemize}
    \item \texttt{plx} (Parallax): The apparent shift in a star’s position due to Earth’s orbital motion, measured in milliarcseconds (mas). Inversely proportional to distance ($d = 1/\texttt{plx}$ in arcseconds).
    
    \item \texttt{pmra}, \texttt{pmdec} (Proper Motion in RA/Dec): The motion of a star across the sky in right ascension and declination directions, respectively, measured in mas/yr. Reflects tangential velocity projected onto the sky.
    
    \item \texttt{Teff} (Effective Temperature): The surface temperature of a star assuming it behaves like a blackbody, expressed in Kelvin. Influences stellar color and spectral type.
    
    \item \texttt{rad} (Radius): The stellar radius, typically in solar radii ($R_\odot$). Derived from luminosity and temperature using Stefan–Boltzmann law.
    
    \item \texttt{mass} (Mass): The estimated mass of the star, usually in solar masses ($M_\odot$), often inferred from stellar models.
    
    \item \texttt{lum} (Luminosity): Total energy output of the star per unit time, expressed in solar luminosities ($L_\odot$). Strongly depends on both radius and temperature.
    
    \item \texttt{logg} (Surface Gravity): Logarithm of the surface gravity ($g$) of the star, in cgs units. Sensitive to both mass and radius; higher for compact stars.
    
    \item \( v_{\text{tan}} \) (Tangential Velocity): The transverse velocity of the star relative to the Sun, calculated from proper motion and distance. Expressed in km/s.
    
    \item \texttt{ruwe} (Renormalized Unit Weight Error): A goodness-of-fit statistic from Gaia astrometry; elevated values often indicate binarity, excess motion, or unresolved sources. Used as a quality indicator for Gaia solutions. \end{itemize}

\section{Tables and Figures}

\begin{figure*}
    \centering
    \includegraphics[width=1\linewidth]{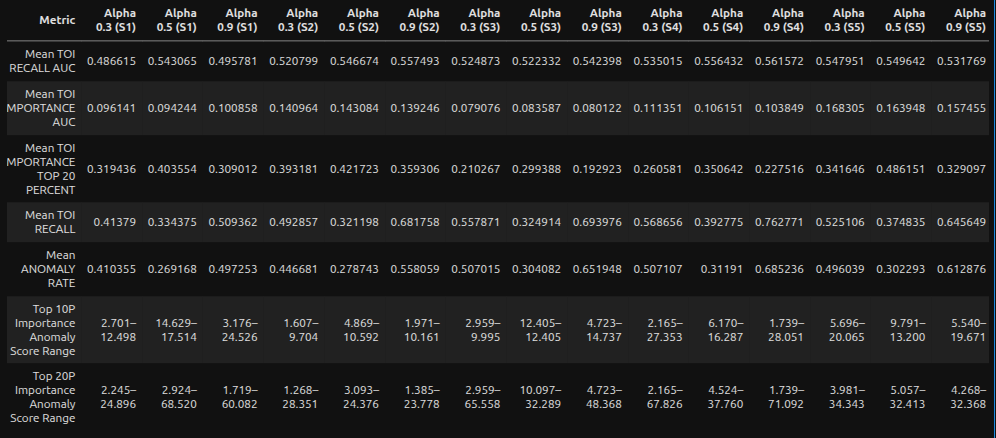}
    \caption{Alpha variants cross sector test performance metrics values}
    \label{fig:alpha-variants-cross-sector-test-performance-metrics-values}
\end{figure*}

\begin{table}[]
    \centering

\caption{Performance comparison of URF variants on TOI detection. TP = True Positives, FP = False Positives, TN = True Negatives, with precision and recall calculated for the 175 total TOIs in the test set for URFs 1,2 and 4 and 195 total TOIs in the test set for URF 3}
\label{tab:urf_benchmarkmodels_performance}
\begin{tabular}{lcccccc}
\hline
Variant & TP & FP & TP+FP & TN & Precision& Recall\\
\hline
URF-1 & 22 & 2159 & 2181 & 0 & 1.01 & 12.57 \\
URF-2 & 45 & 3990 & 4035 & 0 & 1.12 & 25.71 \\
URF-3 & 14 & 300 & 314 & 15499 & 4.46 & 7.18 \\
URF-4 & 64 & 4982 & 5046 & 0 & 1.27 & 36.57 \\
\hline
\end{tabular}

\end{table}

\begin{table}
\centering
\caption{Initial injection--recovery test results at threshold = 0.35. 
Each configuration consisted of 20 injected and 1980 real TESS light curves.}
\label{tab:injrec_thr035}
\begin{tabular}{lccc}
\toprule
Config & WRSS Retrieved & Curves to Vet & Recovery Ratio \\
\midrule
C1  & 20 & 119 & 0.168 \\
C2  & 20 & 125 & 0.160 \\
C3  & 20 & 121 & 0.165 \\
C4  & 20 & 122 & 0.164 \\
C5  & 20 & 120 & 0.167 \\
C6  & 20 & 130 & 0.154 \\
C7  & 19 & 124 & 0.153 \\
C8  & 19 & 136 & 0.140 \\
C9  & 19 & 120 & 0.158 \\
C10 & 19 & 142 & 0.134 \\
\end{tabular}
\end{table}

\begin{table}
\centering
\caption{Initial injection--recovery test results at threshold = 0.60. 
Each configuration consisted of 20 injected and 1980 real TESS light curves.}
\label{tab:injrec_thr060}
\begin{tabular}{lccc}
\toprule
Config & WRSS Retrieved & Curves to Vet & Recovery Ratio \\
\midrule
C1  & 17 & 22 & 0.773 \\
C2  & 16 & 24 & 0.667 \\
C3  & 19 & 24 & 0.792 \\
C4  & 18 & 25 & 0.720 \\
C5  & 18 & 23 & 0.783 \\
C6  & 14 & 23 & 0.609 \\
C7  & 16 & 23 & 0.696 \\
C8  & 17 & 27 & 0.630 \\
C9  & 16 & 21 & 0.762 \\
C10 & 16 & 31 & 0.516 
\end{tabular}
\end{table}

\begin{table}
\centering
\caption{Initial injection--recovery test results at threshold = 0.90. 
Each configuration consisted of 20 injected and 1980 real TESS light curves.}
\label{tab:injrec_thr090}
\begin{tabular}{lccc}
\toprule
Config & WRSS Retrieved & Curves to Vet & Recovery Ratio \\
\midrule
C1  & 8 & 8 & 1.00 \\
C2  & 4 & 4 & 1.00 \\
C3  & 8 & 8 & 1.00 \\
C4  & 4 & 4 & 1.00 \\
C5  & 10 & 10 & 1.00 \\
C6  & 2 & 2 & 1.00 \\
C7  & 6 & 6 & 1.00 \\
C8  & 4 & 4 & 1.00 \\
C9  & 3 & 3 & 1.00 \\
C10 & 4 & 4 & 1.00 \\
\end{tabular}
\end{table}

\begin{figure}
    \centering
    \includegraphics[width=1\linewidth]{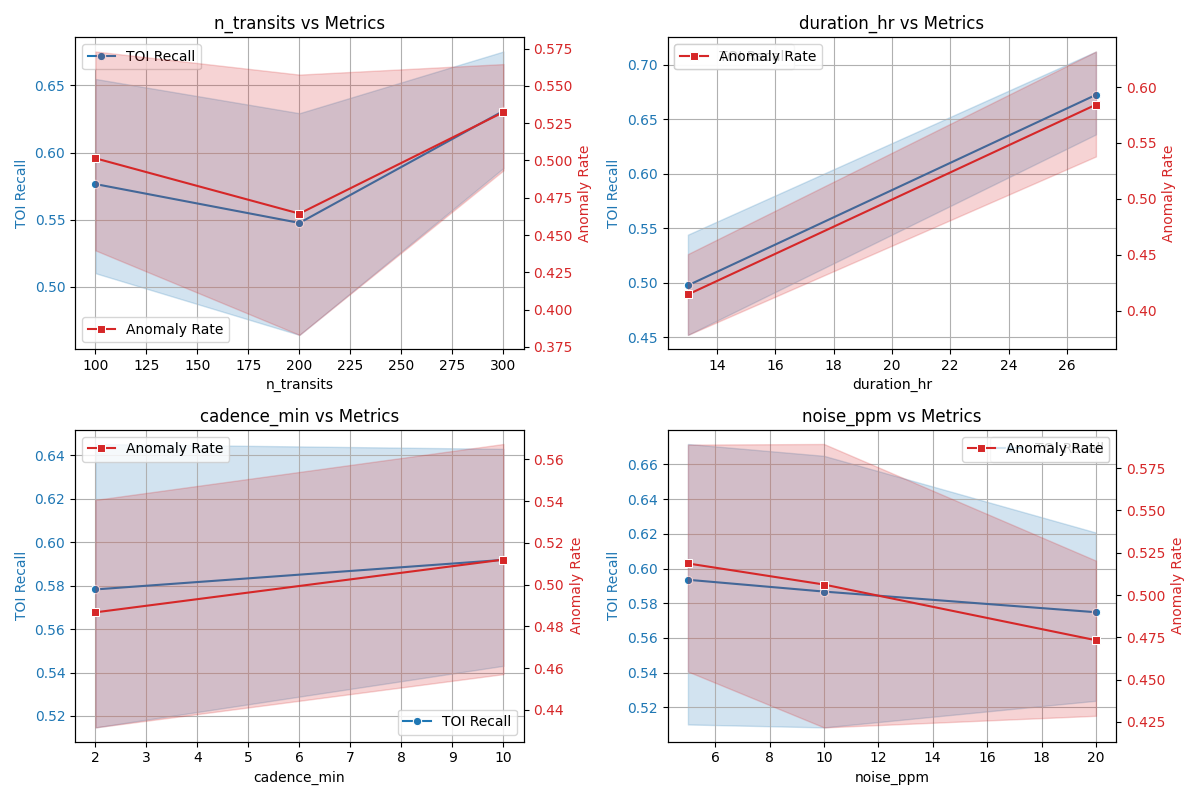}
    \caption{ncurves, duration\_hr (days \ref{sec:noteonduration}), noise (ppm) and cadence (minutes) clockwise from top left against TOI Recall (blue) and Anomaly Rate (red)}
    \label{fig:synthparamsvsrecallanomalyrate1}
\end{figure}

\begin{figure}
    \centering
    \includegraphics[width=1\linewidth]{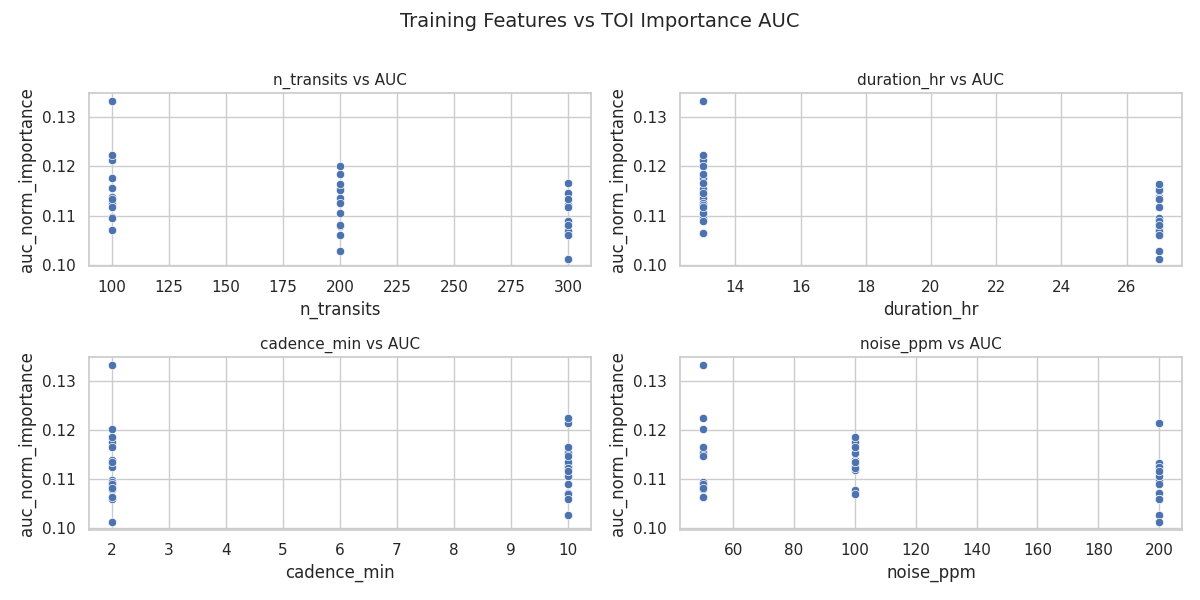}
    \caption{Synthetic features (ncurves, duration\_hr(in days \ref{sec:noteonduration}), noise (ppm), cadence (minutes) - clockwise from top left) v.s TOI Importance AUC}
    \label{fig:synthfeaturesvstoiimportanceauc}
\end{figure}

\begin{figure}
    \centering
    \includegraphics[width=1\linewidth]{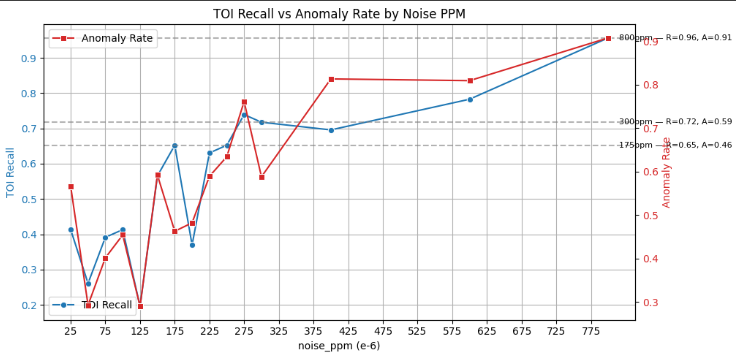}
    \caption{TOI Recall (blue) vs Anomaly Rate (red) by Noise (ppm) with values [25, 50, 75, 100, 125, 150, 175, 200, 225,
250, 275, 300, 400, 600, 800]}
    \label{fig:recallvsanomalyratebynoisefinermapping}
\end{figure}

\begin{figure}
    \centering
    \includegraphics[width=1\linewidth]{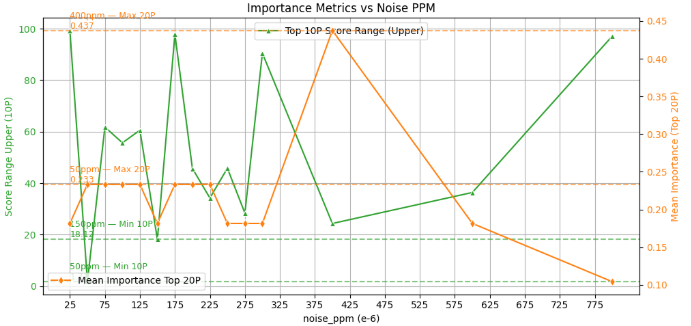}
    \caption{Mean Importance Top 20P and Anomaly Score Range Upper 10P vs Noise (ppm)}
    \label{fig:importancevsnoisefinermapping}
\end{figure}

\begin{figure}
    \centering
    \includegraphics[width=1\linewidth]{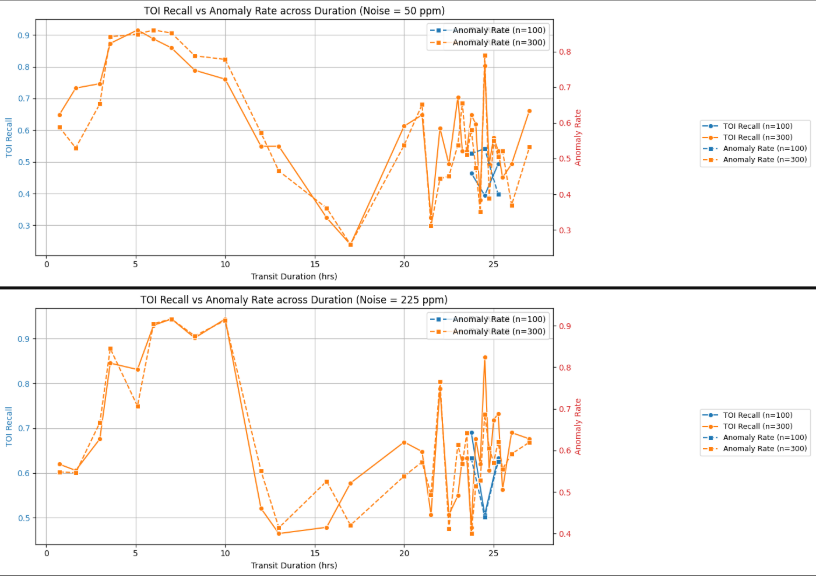}
    \caption{Recall vs Anomaly Rate across Duration values for Noise = 50, 225. Yellow line is  for n\_curves=300, blue for n\_curves=100}
    \label{fig:durationvariationfinermapping}
\end{figure}

\begin{figure}
    \centering
    \includegraphics[width=1\linewidth]{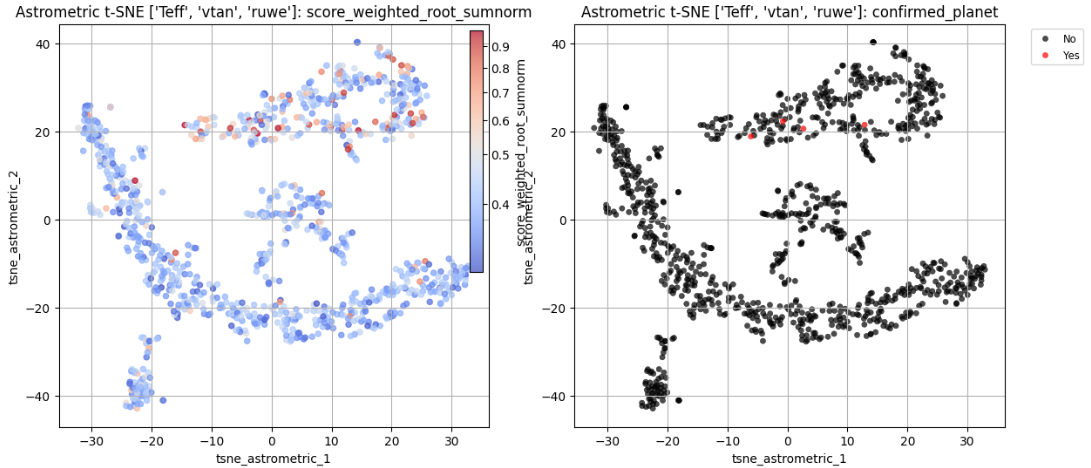}
    \caption{Overlay of confirmed planets (in red on the right) from TESS SPOC Sector 1 in the high N-WRSS cluster}
    \label{fig:confirmedplanetssec1tsneoverlay}
\end{figure}

\begin{table*}
\centering
\caption{Data on the 5 curves shown in appendix from planet-like class of sector 1}
\label{tab:clara_planets}
\begin{tabular}{lllllll }
\hline
TIC ID & Planet Name & Period (days) & Mass ($M_\mathrm{J}$) & Radius ($R_\mathrm{J}$) & Stellar Type & Discovery Status \\
\hline
149603524 & WASP-62b & $4.411953 \pm 3\times10^{-6}$ & -- & $1.39 \pm 0.06$ & G-type & Confirmed \\
144065872 & WASP-95b & $2.18466 \pm 2\times10^{-5}$ & 1.44 & $1.23 \pm 0.06$ & G-type & Confirmed (2014) \\
260609205 & TESS Candidate & 4.4616249 & -- & 2.83 & Evolved (1.19 $M_\odot$) & TESS TCE \\
290131778 & HD 202772A b & 3.3 & $1.017^{+0.070}_{-0.068}$ & $1.545^{+0.052}_{-0.060}$ & Mildly evolved & Confirmed \\
92352620 & -- & 2.0 & 0.618 & 1.26 (est.) & -- & Confirmed (2014) \\
\hline
\end{tabular}
\end{table*}

\begin{table}
\centering
\caption{SIMBAD label frequencies for the weakly supervised anomaly subset retrieved via MAST coordinate cross-matching. Labels are grouped into broader astrophysical categories for downstream classification.}
\label{tab:simbad_label_freq}
\begin{tabular}{lcl}
\hline
\textbf{SIMBAD Label} & \textbf{Count} & \textbf{Label Group} \\
\hline
*      & 230 & Stellar \\
Pl?    & 119 & Planet-like \\
PM*    & 113 & Stellar \\
err    & 79  & Outlier \\
Pl     & 68  & Planet-like \\
SB*    & 23  & Binary Star \\
**     & 6   & Binary Star \\
EB*    & 6   & Binary Star \\
Er*    & 2   & Stellar \\
s*b    & 2   & Planet-like \\
LM*    & 1   & Binary Star \\
BD*    & 1   & Planet-like \\
\hline
\end{tabular}
\end{table}

\begin{table}[H]
\centering
\caption{TOIs in DPMM Cluster 2}
\label{tab:dpmmc2tois}
\begin{tabular}{lll}
\hline
\textbf{\#} & \textbf{TIC ID} & \textbf{TOI} \\
\hline
1  & 92352620  & 107.01 \\
2  & 144065872 & 105.01 \\
3  & 149603524 & 102.01 \\
4  & 260609205 & 219.01 \\
5  & 38846515  & 106.01 \\
6  & 149603524 & 102.01 \\
7  & 184240683 & 250.01 \\
8  & 230982885 & 195.01 \\
9  & 388104525 & 112.01 \\
10 & 38846515  & 106.01 \\
11 & 260609205 & 219.01 \\
12 & 349518800 & 407.01 \\
13 & 388104525 & 112.01 \\
14 & 122612091 & 264.01 \\
15 & 38686737  & 432.01 \\
16 & 38846515  & 106.01 \\
\hline
\end{tabular}
\end{table}

\begin{table}[H]
\centering
\caption{TOIs in GMM Cluster 0}
\label{tab:gmmc0tois}
\begin{tabular}{ll}
\hline
\textbf{\#} & \textbf{TIC ID} \\
\hline
1  & 92352620 \\
2  & 144065872 \\
3  & 149603524 \\
4  & 260609205 \\
5  & 38846515 \\
6  & 149603524 \\
7  & 184240683 \\
8  & 230982885 \\
9  & 388104525 \\
10 & 38846515 \\
11 & 100100827 \\
12 & 260609205 \\
13 & 349518800 \\
14 & 388104525 \\
15 & 122612091 \\
16 & 38686737 \\
17 & 38846515 \\
\hline
\end{tabular}
\end{table}

\begin{table}
\centering
\caption{Configuration ranges for the pool of 200 synthetic injected transit curves. Each configuration contributes 20 curves, with depth ranges expressed in parts per million (ppm) and white noise levels specified in ppm.}
\label{tab:mc_configs}
\begin{tabular}{lccc}
\hline
Config ID & Depth Range (ppm) & Noise (ppm) & \# Curves \\
\hline
C1  & 200–500     & 50   & 20 \\
C2  & 200–500     & 200  & 20 \\
C3  & 500–1000    & 50   & 20 \\
C4  & 500–1000    & 300  & 20 \\
C5  & 1000–2000   & 100  & 20 \\
C6  & 1000–2000   & 400  & 20 \\
C7  & 2000–5000   & 150  & 20 \\
C8  & 2000–5000   & 500  & 20 \\
C9  & 5000–10000  & 300  & 20 \\
C10 & 10000–20000 & 600  & 20 \\
\hline
\end{tabular}
\end{table}

\begin{figure}
    \centering
    \includegraphics[width=1\linewidth]{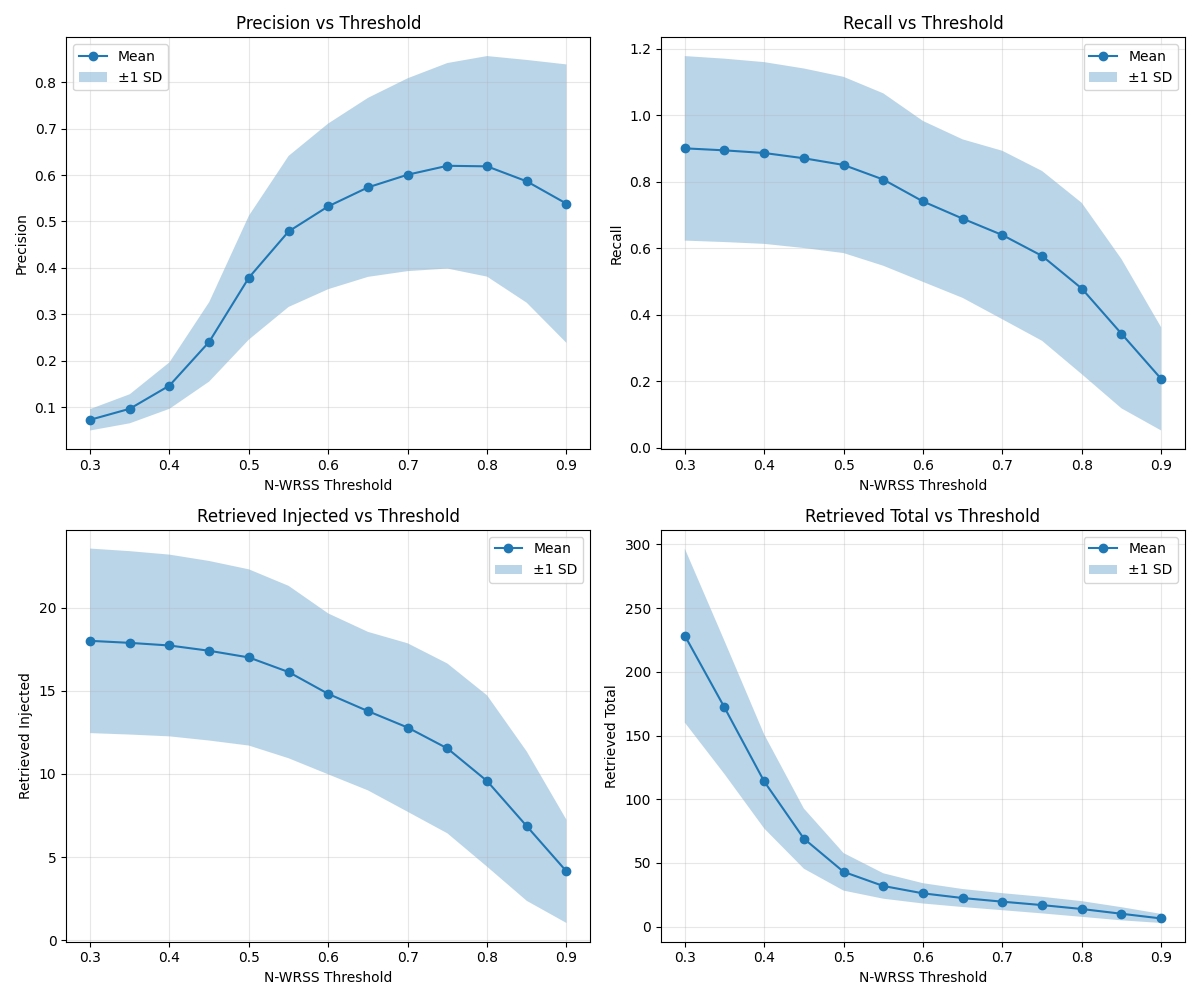}
    \caption{Monte Carlo Injection-Recovery performance metrics.
        Each panel shows the mean (solid line) and ±1 standard deviation (shaded area) for (a) Precision, 
        (b) Recall, (c) Number of injected signals retrieved, and (d) Total number of signals retrieved. 
        The results are based on Monte Carlo injection-recovery tests, illustrating how threshold choice affects detection performance.}
    \label{fig:mc_injrec_thresholdsweep}
\end{figure}

\begin{figure}
    \centering
    \includegraphics[width=1\linewidth]{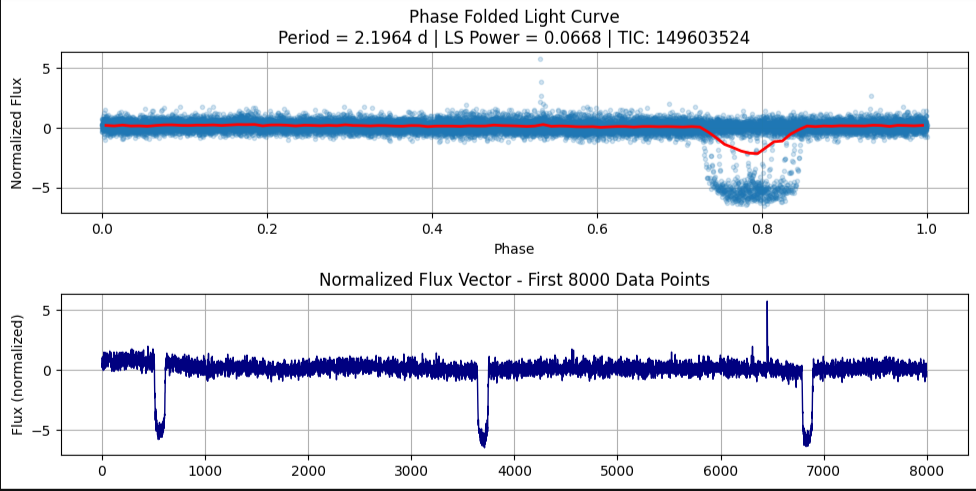}
\end{figure}

\begin{figure}
    \centering
    \includegraphics[width=1\linewidth]{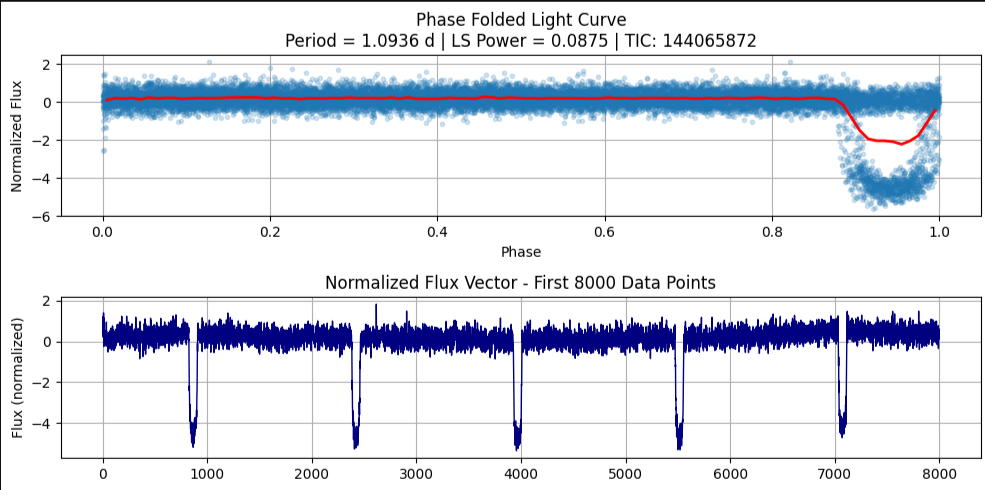}
\end{figure}
\begin{figure}
    \centering
    \includegraphics[width=1\linewidth]{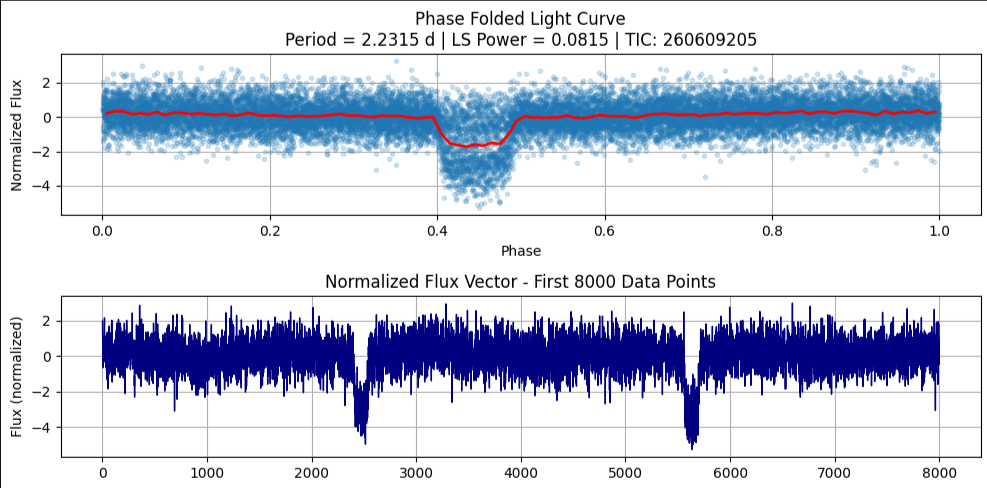}
\end{figure}

\begin{figure}
    \centering
    \includegraphics[width=1\linewidth]{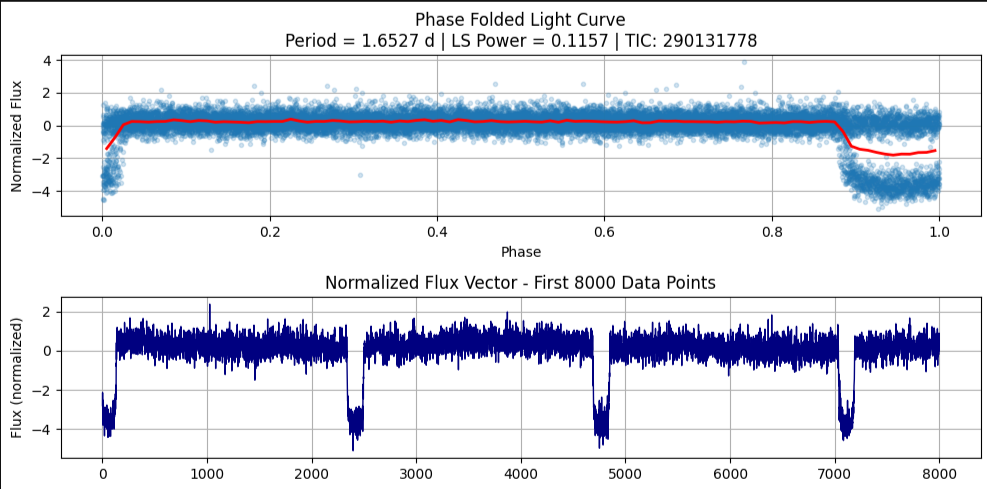}
\end{figure}

\begin{figure}
    \centering
    \includegraphics[width=1\linewidth]{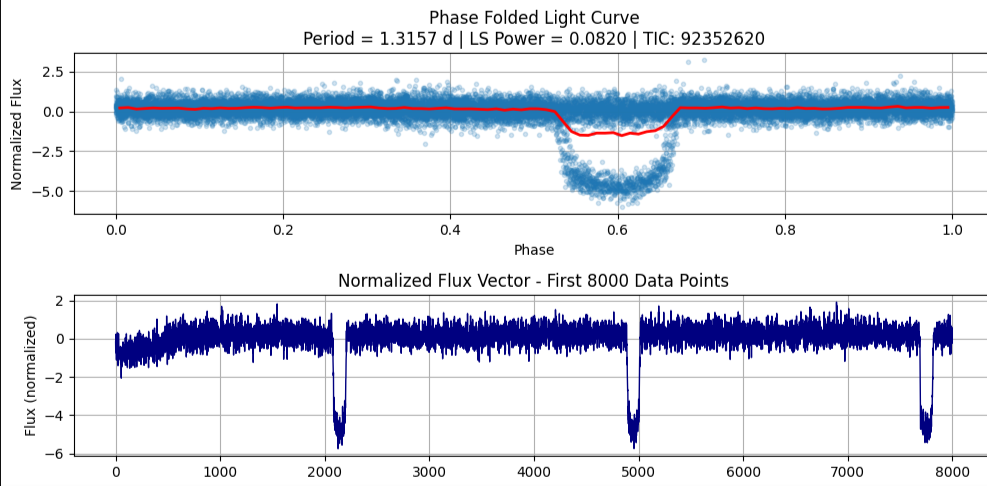}
\end{figure}



\end{document}